\pdfoutput=1
\documentclass[aps,prd,amsmath,floats,floatfix,twocolumn,
superscriptaddress,nofootinbib,showpacs,longbibliography]{revtex4-1}

\usepackage{packages}

\begin{document}

\title{Searching for additional structure and redshift evolution in the observed binary black hole population with a parametric time-dependent mass distribution}

\newcommand{\ldit}{\affiliation{Laboratoire des 2 Infinis - Toulouse (L2IT-IN2P3), Université de Toulouse, CNRS, F-31062 Toulouse Cedex 9, France}}
\newcommand{\INFN}{\affiliation{INFN, Sezione di Roma, I-00185 Roma, Italy}}
\newcommand{\ipdi}{\affiliation{Université Lyon, Université Claude Bernard Lyon 1, CNRS, IP2I Lyon/IN2P3,UMR 5822, F-69622 Villeurbanne, France}}

\author{Vasco Gennari}\ldit
\author{Simone Mastrogiovanni}\INFN
\author{Nicola Tamanini}\ldit
\author{Sylvain Marsat}\ldit
\author{Gregoire Pierra}\ipdi
\email{vasco.gennari@l2it.in2p3.fr}

\hypersetup{pdfauthor={Gennari et al.}}

\date{\today}

\begin{abstract}
The population of the observed gravitational wave events encodes unique information on the formation and evolution of stellar-mass black holes, from the underlying astrophysical processes to the large-scale dynamics of the Universe.
We use the \texttt{ICAROGW} analysis infrastructure to perform hierarchical Bayesian inference on the gravitational wave signals from the LIGO-Virgo-KAGRA third observing run, O3.
Searching for additional structure and redshift evolution in the primary mass distribution, we explore the dependence of the mass spectrum reconstruction on different parametrizations and prior choices.
For the stationary case, we find strong evidence (Bayes factor $\mathcal{B} \simeq 180$) that the results obtained using a power-law model with a peak (\model{Powerlaw-Gaussian})--the model preferred so far in the literature--are sensitive to prior bounds, affecting the resolvability of the $\sim 35 M_{\odot}$ peak.
This behaviour is reproduced by simulated data, indicating a bimodal structure in the likelihood.
Models with three mass features simultaneously capture a sharp $\sim 10M_{\odot}$ peak, a $\sim 35 M_{\odot}$ overdensity, and support for a $\sim 20 M_{\odot}$ overdensity preceded by a dip.
Among these, a model with three power-law peaks (\model{Powerlaw-Powerlaw-Powerlaw}) is equally favored, in terms of evidence, over the \model{Powerlaw-Gaussian} model with wide priors.
We find no statistical support for redshift evolution in the current data and provide constraints on the parameters governing this evolution, showing consistency with stationarity.
We highlight possible limitations of the hierarchical Bayesian inference framework in reconstructing evolving features outside the detector horizon.
Our work lays the foundations for a robust characterization of time-dependent population distributions, with significant implications for black hole astrophysics and gravitational wave cosmology.
\end{abstract}

\maketitle
\tableofcontents

\section{Introduction}
\label{sec:introduction}

One of the main objectives of gravitational wave (GW) physics is the characterization of the observed binary black hole (BH) population, which contains crucial information on the astrophysical processes governing the formation and evolution of BH mergers~\cite{LIGOScientific:2020kqk, KAGRA:2021duu, Wysocki:2018mpo, Belczynski:2017gds, Farr:2017uvj, OShaughnessy:2017eks, Fishbach:2017dwv, Doctor:2019ruh, McKernan:2019beu, Barrett:2017fcw, Farr:2017gtv, Gerosa:2018wbw, Mandel:2020cig, Romero-Shaw:2020siz, Stevenson:2015bqa, Tiwari:2018qch, Wong:2020ise, Wysocki:2017isg, Gerosa:2021mno, Zevin:2020gbd}.
Reconstructing such a distribution allows, in principle, to distinguish between different binary BH formation channels~\cite{Zevin:2017evb, Mandel:2018hfr, Zevin:2020gbd, Mapelli:2020vfa, Mapelli:2021taw, Marchant:2023wno, Baibhav:2020xdf, Gerosa:2017kvu, Mapelli:2021syv, Antonini:2018auk, Kimball:2020opk, Tagawa:2021ofj, Mapelli:2020xeq, Gerosa:2021mno}, and to infer cosmology using only gravitational information~\cite{Markovic:1993cr, Chernoff:1993th, Finn:1995ah, Taylor:2011fs, Farr:2019twy, Ezquiaga:2022zkx, Mastrogiovanni:2022hil, LIGOScientific:2019zcs, LIGOScientific:2021aug}.
In Bayesian analysis, this problem is typically addressed through a \textit{hierarchical approach}, where the posterior probability distribution of the inferred population is the preferred one from which single events parameters are drawn, simultaneously accounting for selection effects induced by the finite detector sensitivity~\cite{Loredo:2001rx, Mandel:2018mve, Gaebel:2018poe, Kapadia:2019uut, Vitale:2020aaz, Gair:2022zsa, Thrane:2018qnx}.
Various methods have been developed for this purpose.
The \textit{parametric} approach, where the population distribution is parametrized in terms of phenomenological models, developed to capture specific features in the data~\cite{Talbot:2018cva, Wysocki:2018mpo, Fishbach:2018edt}.
In this framework, different competing models can be compared through \textit{Bayesian model selection}, for example by computing the Bayes factors (BFs)~\cite{LIGOScientific:2020kqk, KAGRA:2021duu}.
Rather than committing to specific parametrizations, more flexible \textit{semi-parametric} and \textit{non-parametric} methods have also been explored, which allow the distribution to be reconstructed from the data with fewer assumptions about the model, and to capture possible unexpected features in the population~\cite{Mandel:2016prl, Tiwari:2020vym, Tiwari:2020otp, Edelman:2021zkw, Tiwari:2021yvr, Sadiq:2021fin, Ruhe:2022ddi, Toubiana:2023egi, Ray:2023upk, Farah:2023vsc, Callister:2023tgi, Heinzel:2023hlb, Heinzel:2024jlc, Heinzel:2024hva, Rinaldi:2023bbd}.

Current population studies mostly focus on characterizing the parameters subspace given by the BH masses, spins and luminosity distance.
Since the first GW detection~\cite{LIGOScientific:2016aoc}, much has been learned about their observed distribution~\cite{Callister:2024cdx}, such as the presence of substructures in the primary mass function at $\sim 10 M_{\odot}$ and $\sim 35 M_{\odot}$~\cite{LIGOScientific:2020kqk, KAGRA:2021duu, Tiwari:2020otp, Tiwari:2020vym, Tiwari:2021yvr, Toubiana:2023egi, Farah:2023vsc, MaganaHernandez:2024qkz}, the evolution of the binary merger rate with redshift~\cite{LIGOScientific:2020kqk, KAGRA:2021duu, Fishbach:2018edt, vanSon:2021zpk, Turbang:2023tjk, Schiebelbein-Zwack:2024roj}, the absence of a sharp cut-off at high masses~\cite{LIGOScientific:2020kqk, KAGRA:2021duu, Edelman:2021fik}, the preference for small effective spins~\cite{LIGOScientific:2020kqk, KAGRA:2021duu, Tiwari:2018qch, Roulet:2021hcu, Hoy:2021rfv}.
In addition, population-level correlations between different parameters have been extensively studied, providing evidence for BH effective spin to increase with unequal mass ratios~\cite{Callister:2021fpo, Adamcewicz:2022hce, KAGRA:2021duu, Heinzel:2023hlb}, the effective spin distribution broadening with redshift~\cite{Bavera:2022mef, Biscoveanu:2022qac, Heinzel:2023hlb}, high-mass BHs having larger spins~\cite{Tiwari:2021yvr, Franciolini:2022iaa, Biscoveanu:2022qac, Li:2023yyt, Godfrey:2023oxb, Ray:2024hos, Pierra:2024fbl, Antonini:2024het}.
However, a common simplification is to assume that the masses are uncorrelated with redshift, and that all the information about time evolution is contained only in the event rate, which is equivalent to considering the mass function to be stationary with time.
Non-parametric studies have been used to search for mass-redshift correlations~\cite{Ray:2023upk, Rinaldi:2023bbd, Heinzel:2024hva, Sadiq:2025aog}, even providing evidence for the evolution of the primary distribution to larger masses for higher redshifts~\cite{Rinaldi:2023bbd}.
Previous parametric analyses have also adopted different strategies to incorporate redshift-dependent parametrizations, mostly focusing on the evolution of the high-mass spectrum~\cite{Fishbach:2021yvy}, correlations of mass with rate evolution~\cite{vanSon:2021zpk}, evolution of the $\sim 35 M_{\odot}$ overdensity traced by metallicity~\cite{Karathanasis:2022rtr}, but never tackling a more general and systematic description of the full redshift dependence on current data.
If a time evolution of the mass function is expected from various astrophysical scenarios~\cite{Belczynski:2010tb, Rodriguez:2019huv, Safarzadeh:2019ctv, Vink:2020nak, vanSon:2021zpk, Karathanasis:2022rtr, Torniamenti:2024uxl, Ye:2024ypm}, its observational characterization is fundamental to control systematics in the population inference, with potentially important consequences for both astrophysics and cosmology with GWs~\cite{Mukherjee:2021rtw, Pierra:2023deu, Agarwal:2024hld}.
Recently, a study searched for the redshift evolution of the primary features~\cite{Lalleman:2025xcs}, following an analysis similar in spirit to the one developed here.
Although our conclusions are similar, our systematic investigation complements the previous study by adding more flexible models and by investigating the dependence on population priors.

In this work, we rely on a primary mass function that accounts for redshift evolution to analyse the GW events from the third LIGO-Virgo-KAGRA (LVK) observing run, O3~\cite{LIGOScientific:2020ibl, LIGOScientific:2021djp}.
We use \texttt{ICAROGW}~\cite{Mastrogiovanni:2022hil, Mastrogiovanni:2023emh, Mastrogiovanni:2023zbw}, a \texttt{python} software designed to perform hierarchical inference and cosmological analyses on GW signals.
With increasing model complexity and evolving degrees of freedom, we extensively characterize the time dependence of various features in the primary distribution and investigate the robustness of different assumptions underlying our phenomenological models.
Although our strategy is conceptually similar to that in~\cite{Fishbach:2021yvy, Lalleman:2025xcs}, we consider a larger set of phenomenological models and different redshift parametrizations, and we statistically evaluate our reconstructions through Bayesian model selection.
Lastly, we employ a novel three-feature population model that allow us to reconstruct additional substructures in the BH mass spectrum.

Our results indicate that, assuming a stationary \model{Powerlaw-Gaussian} model for the primary mass, prior bounds-dependent results affect the resolvability of the $\sim 35 M_{\odot}$ peak.
Notably, our analysis with wide priors on the Powerlaw slope $\alpha$ and the Gaussian width $\sigma$ is substantially preferred over the one with narrower priors typically used in the literature, by a BF of $\text{ln}\,\mathcal{B} = 5.2 \pm 0.2$ ($\mathcal{B} \simeq 180$).
We reproduce this finding by performing simulations, which support the existence of a bimodal likelihood.
We also propose models with additional features that are able to retrieve more structure in the data, in marked agreement with previous semi- and non-parametric analyses~\cite{Tiwari:2020otp, Tiwari:2021yvr, Toubiana:2023egi, Edelman:2021zkw, Sadiq:2021fin, Farah:2023vsc, Callister:2023tgi, KAGRA:2021duu}.
Among these models, the \model{Powerlaw-Powerlaw-Powerlaw} gives comparable evidence to the (wide prior) \model{Powerlaw-Gaussian}, yielding a relative BF of $\text{ln}\,\mathcal{B} = 0.3 \pm 0.2$.
%
We find no support for redshift evolution using different parametrizations, with the maximum BF in favour of evolution being $\text{ln}\,\mathcal{B} = -1.2 \pm 0.2$.
Using the preferred \model{Powerlaw-Powerlaw-Powerlaw} model, we constrain the linear redshift evolution of its features, specifically the $\sim 10 M_{\odot}$ peak, and the $\sim 20 M_{\odot}$ and $\sim 35 M_{\odot}$ overdensities.
Our analysis reveals a peculiar interplay between evolving parametrizations and selection effects, as the presence of features outside the detector horizon have a minor effect on the observed distributions.
Although the effect is currently marginal, our results seem to point to potential limitations in reliably reconstructing time-dependent astrophysical populations beyond detector sensitivity.

In Sec.~\ref{sec:model}, we discuss the foundations of our framework, focusing on the derivation of a general parametrization for time-evolving populations.
Sec.~\ref{Sec:simulations} is dedicated to testing our evolving models on simulated populations, which also allows us to investigate the occurrence of systematic biases induced by neglecting evolution in the model in the presence of intrinsically evolving data.
In Sec.~\ref{sec:O3_data_analysis}, we perform a hierarchical population analysis on the GW signals from the third LVK observation run, O3, comparing different parametrizations by Bayesian model selection.
We comment on the results and their implications in Sec.~\ref{sec:discussion}, and conclude in Sec.~\ref{sec:conclusions}.

\section{Phenomenological evolving model}
\label{sec:model}

\subsection{Hierarchical likelihood}

To get information on the population of BH, we need to combine observations of multiple events.
This is achieved by \textit{hierarchical} analysis,
which addresses the question of what is the probability distribution from which the observations are drawn.
In a Bayesian framework, the likelihood adopted for such a formulation is generally different from the single-event one used to describe a GW signal in the presence of noise~\cite{Thrane:2018qnx}, and it depends on the specific problem at hand~\cite{LIGOScientific:2020kqk, KAGRA:2021duu, LIGOScientific:2017adf, LIGOScientific:2019zcs, LIGOScientific:2021aug, LIGOScientific:2021aug, LIGOScientific:2016lio}.
In the case of the observed binary mergers, the likelihood is normally modelled with an \textit{inhomogeneous Poisson process}, that captures the information on the event rate in time.
Under these assumptions, the following GW hierarchical likelihood $\mathcal{L}_{H}$ can be derived~\cite{Mandel:2018mve, Vitale:2020aaz}
\begin{multline}\label{hierarchical_likelihood}
    \mathcal{L}_{H}\qty(\{\vb{x}\}|\vb{\Lambda}) \propto e^{-N_{\text{exp}}(\vb{\Lambda})}\: \prod_{i}^{N_{\text{gw}}} T_{\text{obs}}\\ \int \dd\bs{\theta}\dd z\: \mathcal{L}_{\text{gw}}\qty(\vb{x}_i\,|\,\bs{\theta}, \vb{\Lambda})\:\frac{\dd \mathcal{N}}{\dd\bs{\theta}\dd z\dd t_d}(\vb{\Lambda}),
\end{multline}
where $\bs{\theta}$ are the \textit{intrinsic} single-event parameters describing the binary, $\vb{\Lambda}$ the \textit{population parameters}, $z$ is the \textit{redshift}, $\{\vb{x}\}$ the set of GW observations, normally represented by the ensemble of samples obtained from the parameter estimation (PE) of each event (e.g.~\cite{LIGOScientific:2021djp}).
While the product over the GW events in the likelihood combines the information from multiple observations, the expected number of events $N_{\text{exp}}$ accounts for the finiteness of the detector sensitivity, usually referred to as \textit{selection effects}~\cite{Loredo:2001rx, Mandel:2018mve, Vitale:2020aaz}.
Thus, in this picture, the single-event likelihood $\mathcal{L}_{\text{gw}}$ represents the ``data'', and the event rate in the detector frame $\frac{\dd \mathcal{N}}{\dd\bs{\theta}\dd z\dd t_d}$ the parametrization of the population distribution, i.e. the ``model''.
The latter is usually decomposed as
\begin{equation}\label{eq:events_rate}
    \frac{\dd \mathcal{N}}{\dd \bs{\theta} \dd z \dd t_d}(\vb{\Lambda}) = \mathcal{R}(z\,|\, \vb{\Lambda})\: p_{\text{pop}}(\bs{\theta}\,|\,z,\vb{\Lambda})\: \frac{1}{1+z} \frac{\dd V_c}{\dd z},
\end{equation}
where $\mathcal{R}(z) \equiv \frac{\dd \mathcal{N}}{\dd V_c \dd t_s}$ is the rate per comoving volume $V_c$ in the source frame, and $p_{\text{pop}}(\bs{\theta}\,|\,z,\vb{\Lambda})$ the probability density function that models the distribution of GW sources in the single-event parameter space.
This probability is independent of the redshift if the distribution of binary parameters does not evolve with cosmic time.
In the rest of this section, we discuss different ways to parametrize redshift evolution in the population distribution $p_{\text{pop}}$.

\subsection{Evolving mass distribution}
\label{Sec:evolving_factorisation}

We now focus on the population probability distribution $p_{\text{pop}}(\bs{\theta}\,|\,z,\vb{\Lambda})$, where $\bs{\theta}$ is some set of event-level parameters.
Current analyses mostly focus on the intrinsic parameters\footnote{Some work also included the sky position to look for anisotropies in the observed BH distribution~\cite{Stiskalek:2020wbj, Essick:2022slj}.}, given by the BH source \textit{masses} $m_{1,2}$ and \textit{spins} $\vb{\chi}_{1,2}$ (e.g.~\cite{LIGOScientific:2020kqk, KAGRA:2021duu}).
To enhance population features or reduce the dimensionality of the problem, different single-event parametrizations can be considered, such as using the \textit{mass ratio} $q \equiv m_2/m_1$ instead of the secondary mass or using mass-spin combinations such as the effective spin $\chi_{eff}$ and precessing spin $\chi_p$ (see~\cite{Callister:2024cdx} and references therein).
In this work, we neglect all information about the spins and restrict ourselves to the parameters $\qty{m_1, q}$.
Although population-level correlations have already been observed for spins and masses~\cite{Franciolini:2022iaa, Heinzel:2023hlb, Pierra:2024fbl, Antonini:2024het, Li:2023yyt, Godfrey:2023oxb}, spins and redshifts~\cite{Biscoveanu:2022qac, Heinzel:2023hlb, Rinaldi:2023bbd}, we assume here that neglecting such correlations does not change our conclusions, since we are mainly interested in the correlations between the primary mass and the redshift.

Under these assumptions, we can write the joint population distribution as
\begin{align}\label{eq:ppop_explicit}
    p_{\text{pop}}\qty(\bs{\theta}\,|\,z,\vb{\Lambda}) &= p_{\text{pop}}\qty(m_1,q\,|\,z,\vb{\Lambda}) \nonumber\\
    &= p_{\text{pop}}\qty(m_1\,|\,z,\vb{\Lambda})\:p_{\text{pop}}\qty(q\,|\,m_1,z,\vb{\Lambda}) \nonumber\\ 
    &= p_{\text{pop}}\qty(m_1\,|\,z,\vb{\Lambda})\:p_{\text{pop}}\qty(q\,|\,z,\vb{\Lambda}),
\end{align}
where in the last step we have further neglected correlations between the mass ratio and the primary mass (e.g. see~\cite{KAGRA:2021duu}).
Assuming the masses to be uncorrelated with the redshift, Eq.~\eqref{eq:ppop_explicit} simplifies to $p_{\text{pop}}\qty(m_1\,|\,\vb{\Lambda})\:p_{\text{pop}}\qty(q\,|\,\vb{\Lambda})$.
This is physically equivalent to assume that the mass distribution is stationary over cosmic time.
Although such assumption is commonly found in the literature~\cite{LIGOScientific:2020kqk, KAGRA:2021duu}, in this work we are mostly interested in addressing the more general problem of time-evolving distributions.
Nonetheless, we still restrict to the case where the mass ratio is stationary, leaving the redshift correlations only in the primary BH mass,
\begin{equation}\label{eq:ppop_final}
    p_{\text{pop}}\qty(\bs{\theta}\,|\,z,\vb{\Lambda}) = p_{\text{pop}}\qty(m_1\,|\,z,\vb{\Lambda})\:p_{\text{pop}}\qty(q\,|\,\vb{\Lambda}).
\end{equation}
This choice is mainly motivated by the non-parametric results in~\cite{Rinaldi:2023bbd}, although such approximation should be relaxed in future, more general analyses.

\subsection{Time-evolving parametrization}
\label{Sec:evolving_model}

From Eq.~\eqref{eq:ppop_explicit}, in our parametric framework, we now need to specify a functional form for the different contributions.
This step is delicate, not only because the results are affected by the chosen parametrization, but also because a robust theoretical motivation to use specific features is still missing.
This situation is very different from single-event analyses, where accurate waveform models contain information from both theoretical physics and numerical simulations of binary mergers~\cite{LIGOScientific:2018mvr, LIGOScientific:2021usb, LIGOScientific:2020ibl, LIGOScientific:2021djp}.
On the contrary, the BH population distribution is the result of all the astrophysical and cosmological processes that intervene in creating, evolving, and merging the observed BHs.
Even though in principle insights on the underlying astrophysical processes can be incorporated in the models~\cite{Golomb:2023vxm, Talbot:2018cva, Karathanasis:2022rtr, Vijaykumar:2023bgs, Ulrich:2024nez, Cheng:2023ddt}, the large uncertainties in current simulations still prevent a reliable prediction of specific feature~\cite{Torniamenti:2024uxl, Vaccaro:2023cwr, Sgalletta:2024jhw, Cheng:2023ddt}.
As a result, phenomenological models are often adopted to capture features in the population, and more flexible semi- or non-parametric analyses are then used to search for additional unmodelled physics.
Even if Bayesian model selection allows to decide which parametrization is preferred to describe the observations, comparing different models is not always trivial, especially when the parameter space starts to change significantly between different models (see Sec.~\ref{sec:O3_data_analysis}).

Suppose we want to parametrize $n$ sub-populations within the full population distribution.
The most general way to account for this is to divide the total source-frame event rate in $n$ contributions~\cite{Mastrogiovanni:2022ykr,Pierra:2024fbl},
\begin{align}
    \frac{\dd \mathcal{N}}{\dd \bs{\theta}\dd V_c \dd t_s} &= \sum_{i=1}^{n} \frac{\dd \mathcal{N}_i}{\dd \bs{\theta}\dd V_c \dd t_s} \nonumber\\
    &= \sum_{i=1}^{n} \mathcal{R}_i(z\,|\,\vb{\Lambda})\: p_{\text{pop,i}}(\bs{\theta}\,|\,z,\vb{\Lambda}),
\end{align}
where $\mathcal{R}_i(z|\mathbf{\Lambda})$ is the merger rate evolution for a specific sub-population.
By defining the overall rate as $\mathcal{R}(z|\mathbf{\Lambda})=\sum_{i=1}^{n} \mathcal{R}_i(z\,|\,\vb{\Lambda})$, and the mixture fractions
\begin{equation}
    \lambda_i(z\,|\,\mathbf{\Lambda}) \equiv \frac{\mathcal{R}_i(z\,|\,\mathbf{\Lambda})}{\mathcal{R}(z\,|\,\mathbf{\Lambda})}, 
\end{equation}
we can rewrite the event rate as 
\begin{equation}
    \frac{\dd \mathcal{N}}{\dd \bs{\theta}\dd V_c \dd t_s} =  \mathcal{R}(z\,|\,\mathbf{\Lambda})\, \sum_{i=1}^{n} \lambda_i(z\,|\,\mathbf{\Lambda}) p_{\text{pop,i}}(\bs{\theta}\,|\,z,\vb{\Lambda}).
    \label{eq:multipopulation}
\end{equation}
From the definition of the mixture fractions, it follows that that $\lambda_1 + ... + \lambda_n = 1$ at all the redshifts.
Note that the $\lambda_i$ coefficients can be generic functions of redshift, allowing the relative weight of the different contributions to vary with time.
From Eq.~\eqref{eq:multipopulation}, it is clear that, for each sub-population, the redshift dependence to be parametrized enters in:
\begin{itemize}
    \item the \textit{mixture function} $\lambda_i(z\,|\,\vb{\Lambda})$;
    \item the mass function $p_{\text{pop,i}}(\bs{\theta}\,|\,z,\bf{\Lambda})$.
\end{itemize}
Although the results can be generalised immediately, for the sake of readability we will focus on the case $n=2$,
\begin{multline}\label{eq:rate_population}
    \frac{\dd N}{\dd \bs{\theta}\dd V_c \dd t_s} = \mathcal{R}(z\,|\,\vb{\Lambda})\: [ \lambda(z\,|\,\vb{\Lambda})\: p_{\text{pop,1}}(\bs{\theta}\,|\,z,\bf{\Lambda})\\ + \qty(1-\lambda(z\,|\,\vb{\Lambda}))\: p_{\text{pop,2}}(\bs{\theta}\,|\,z,\bf{\Lambda})],
\end{multline}
having defined $\lambda(z) \equiv \lambda_1(z) = 1 -\lambda_2(z)$.
Using the set of single-event parameters of Eq.~\eqref{eq:ppop_explicit}, Eq.~\eqref{eq:rate_population} becomes
\begin{align}
    \frac{\dd N}{\dd \bs{\theta}\dd V_c \dd t_s} &= \{\lambda(z)\: p_{\text{pop,1}}(\bs{\theta}\,|\,z)\nonumber\\
    &+ \qty[1-\lambda(z)]\: p_{\text{pop,2}}(\bs{\theta}\,|\,z)\}\: \mathcal{R}(z)\nonumber\\
    &= \{\lambda(z)\: p_{\text{pop,1}}\qty(m_1\,|\,z)\nonumber\\
    &+ \qty[1-\lambda(z)]\: p_{\text{pop,2}}\qty(m_1\,|\,z)\}
    \mathcal{R}(z)\,p_{\text{pop}}\qty(q),
\end{align}
after further requiring that the (stationary and primary-independent) mass ratio follows the same distribution for the two subpopulations, i.e. $p_{\text{pop,1}}\qty(q) = p_{\text{pop,2}}\qty(q) \equiv p_{\text{pop}}\qty(q)$, and omitting the explicit dependence on the population parameters $\vb{\Lambda}$ for space.
Under these assumptions, we finally obtain the redshift dependent primary mass distribution that is used in this work,
\begin{multline}\label{evolving_primary_mass}
    p_{\text{pop}}\qty(m_1\,|\,z,\vb{\Lambda}) = \lambda(z\,|\,\vb{\Lambda})\: p_{\text{pop,1}}\qty(m_1\,|\,z,\vb{\Lambda})\\ + \qty[1-\lambda(z\,|\,\vb{\Lambda})]\: p_{\text{pop,2}}\qty(m_1\,|\,z,\vb{\Lambda}).
\end{multline}
Eq.~\eqref{evolving_primary_mass} tells us that the total redshift evolution in the primary distribution is given by the following two contributions.
\begin{itemize}
    \item A generic redshift dependence of each of the two populations. This models how the shape of the distribution changes with redshift.
    \item A mixture function between the two subpopulations. This models how the relative weight changes with redshift.
\end{itemize}
At this point, it is only a matter of choosing a functional form for the different contributions.
Usually, each subpopulation is associated with a population ``feature'', parametrized by a specific probability distribution.
For example, a redshift evolving \model{Powerlaw-Gaussian} model corresponds to
\begin{itemize}
    \item $p_{\text{pop,1}}\qty(m_1\,|\,z,\vb{\Lambda}) \sim \text{evolving Powerlaw, PL}(z)$;
    \item $p_{\text{pop,2}}\qty(m_1\,|\,z,\vb{\Lambda}) \sim \text{evolving Gaussian, G}(z)$;
    \item $\lambda(z\,|\,\vb{\Lambda}) \sim \text{linear transition}$,
\end{itemize}
such that $\text{PL}(z) \equiv \text{PL}(\alpha(z),\, m_{\text{min}}(z),\, m_{\text{max}}(z))$ and $\text{G}(z) \equiv \text{G} (\mu(z),\, \sigma(z))$.
Finally, each population parameter can in principle have a generic redshift dependence.
Even if we also consider different parametrizations (see Sec.~\ref{Sec:O3_evolving_models}), we mostly focus on a \textit{linear} redshift expansion,
\begin{equation}\label{eq:linear_expansion}
    \vb{\Lambda}(z) \equiv \vb{\Lambda}_{z_0} + \vb{\Lambda}_{z_1}\,z.
\end{equation}
%
Different parametrizations have also been considered in the literature~\cite{Fishbach:2021yvy, Lalleman:2025xcs}.
A more agnostic model with fully independently evolving subpopulations should be investigated in the future, also including redshift evolution in the secondary mass function.

\section{Model validation}\label{Sec:simulations}

\begin{figure*}
    \hspace*{-0.15cm}\includegraphics[scale=0.9]{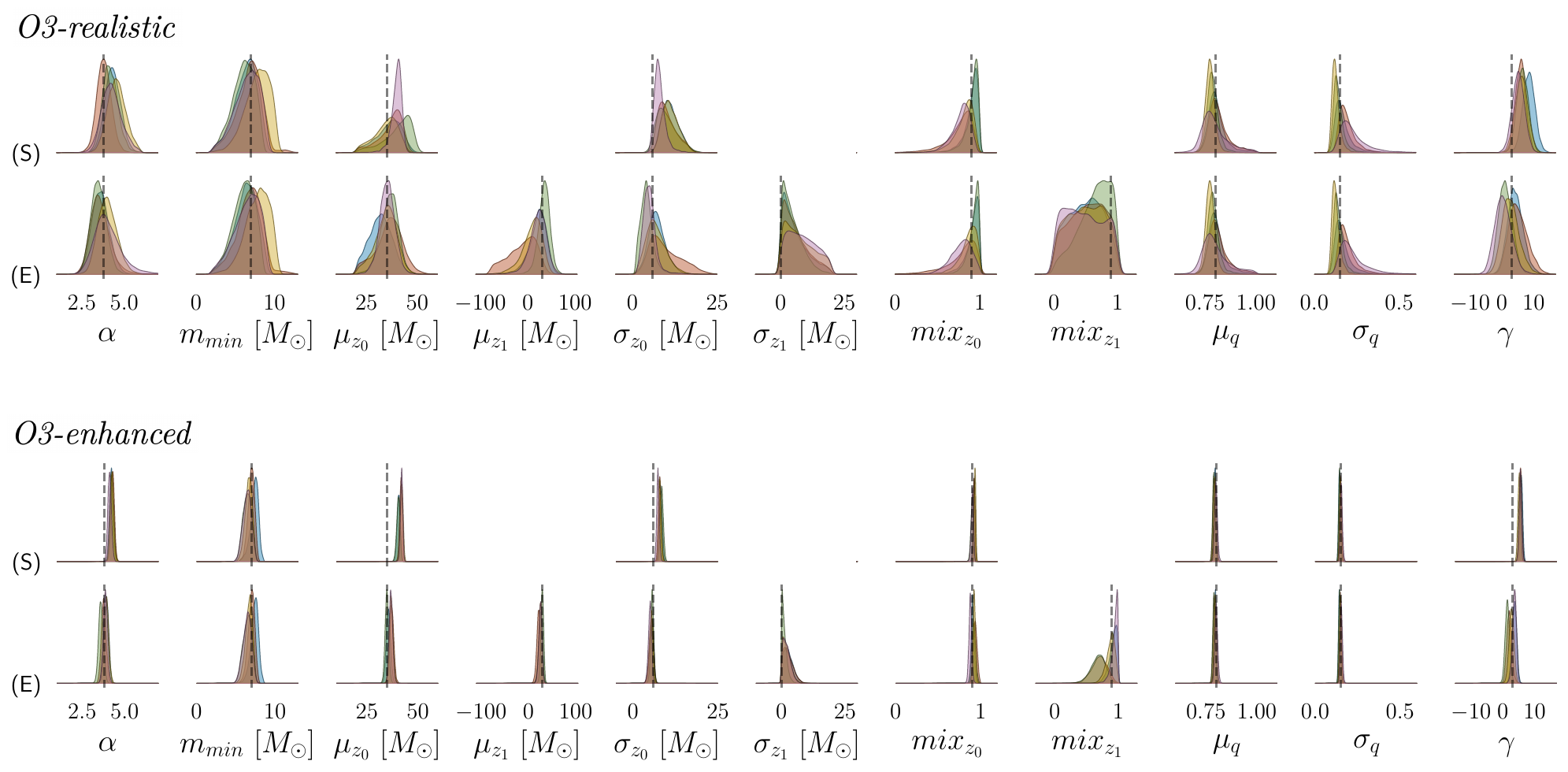}
    \caption{\justifying \footnotesize Posterior distributions of the recovered population parameters with \textit{simulated redshift evolving data}.
    The top (bottom) panel corresponds to the \textit{O3-realistic} (\textit{O3-enhanced}) case with $\sim 50$ ($\sim 1000$) detected events.
    For each case, we show results using a stationary (S) and a redshift evolving (E) model.
    Different colors correspond to five different realizations of the same population.
    The dashed lines correspond to the injected population parameters from which the events are drawn.
    }
    \label{fig:simulation}
\end{figure*}

We validate the redshift evolving model on a simulated population using \texttt{ICAROGW}, a \texttt{python} package for hierarchical inference with GWs~\cite{Mastrogiovanni:2022hil, Mastrogiovanni:2023emh, Mastrogiovanni:2023zbw}.
The events are drawn from a \model{Powerlaw-Gaussian} distribution for the primary BH, first considering a stationary population, then a redshift evolving one.
All the details on the simulation settings are reported in App.~\ref{App:simulations}, and we closely follow the analysis in~\cite{Pierra:2023deu, Pierra:2024wrk}.
In the evolving case, we restrict ourselves to the scenario in which only the Gaussian peak evolves linearly with redshift, and the overall rate of events evolves as a \model{Powerlaw}.
We recover the two populations using a \model{Powerlaw-Gaussian} model, first with a model where the only redshift dependence is in the event rate but the primary distribution is stationary (S), then including a free linear evolution of the Gaussian peak (E).
We study all these combinations in injection and recovery.
While the case of using the same recovery model as the injected one is necessary to validate the model in both the stationary and the evolving case, the case of analyzing an evolving population with a stationary model represents a study of systematics.

We repeat the same analysis for an ``O3-realistic'' scenario, with a $\sim 50$ number of detected events, and an ``O3-enhanced'' one with $\sim 1000$ events.
This allows for a qualitative understanding of how results vary with the number of events, and highlights possible systematics in the analysis that may be harder to expose with fewer events and broader posteriors.
Additionally, to optimise computational resources for our validation, we assume a perfect measurement of the single-event parameters, i.e. for each event we have only one PE sample corresponding to the true event parameters.
This unrealistic but necessary simplification increases the resolution of the analysis by removing all uncertainty associated with the data, while still remaining consistent with our likelihood formulation~\cite{Essick:2023upv}.

Injections to evaluate selection effects are also generated from a stationary \model{Powerlaw-Gaussian} distribution, making sure that the number of injections is sufficient to ensure numerical stability in the likelihood evaluation, even for the evolving scenarios~\cite{Mastrogiovanni:2023zbw, Farr:2019rap}.
We select detected events based on the (matched-filter) \textit{signal-to-noise ratio} (SNR), and consider events to be detected if $\text{SNR}\geq 12$.
We use an analytical approximant to estimate the (single-detector) optimal SNR of the synthetic population and include the effect of a Gaussian noise on the detection, following Sec. II.A of~\cite{Pierra:2023deu} (see also App.~\ref{App:simulations}).
This procedure is not intended to be representative of a proper search in real detector noise, but only to reproduce qualitatively the current LVK response: the results will be consistent as soon as the same procedure is used to select both the events and the injections~\cite{Essick:2023upv}.

The results are summarised in Fig.~\ref{fig:simulation}, where we show the posterior distributions of most of the population parameters for different combinations of evolution and stationarity in recovery.
Note that, for each model, the same analysis is repeated multiple times (different colors in Fig.~\ref{fig:simulation}) to mitigate over the Poisson noise coming from having drawn only a discrete finite set of events from the population.
This effect propagates in our dataset, resulting in a statistical fluctuation of the posterior distributions around the true value.
This is observationally relevant, meaning that with only one draw we should not necessarily expect the posteriors to contain the injected value, especially with a large number of population parameters, as it is the case for some posteriors in Fig.~\ref{fig:simulation}.
However, when the measurement is averaged over several population realizations, it is clear that the posteriors do indeed fluctuate around the injected value.

Under these considerations, we can see that the injected value is correctly recovered if we use the evolving model (E) to analyse the evolving data.
However, as expected, recovering an evolving population with a stationary model (S) leads to a biased parameter estimation, showing that the stationary model is not flexible enough to capture evolving features in the data~\cite{Pierra:2023deu}.
Even if in the O3-realistic case this effect is partially masked by the broader posterior distributions, allowing the true value to fit into the posteriors $90\%$ credible intervals for most of the realizations, it is clear in the O3-enhanced case how the results are systematically biased for some parameters.

\section{LVK O3 data}
\label{sec:O3_data_analysis}

We now analyse real data from the third LVK observing run O3~\cite{LIGOScientific:2020ibl, LIGOScientific:2021djp}, focusing on the binary BH mergers.
To evaluate selection effects, we employ real-noise injections calibrated to O3 sensitivity~\cite{O3:bbhpop, KAGRA:2021duu}, and accordingly use only events from O3 instead of the full GWTC-3 catalog.
To minimise the presence of false triggers in our dataset, we impose a tight threshold for detection by the \textit{Inverse False Alarm Rate} of $\text{IFAR} > 4\, yr$.
This is similar to the latest LVK cosmological analysis~\cite{LIGOScientific:2021aug}, but is a more stringent cut compared to the LVK population study on BHs~\cite{KAGRA:2021duu} (see App.~\ref{sec:dataset} for more information).
Additionally, we do not consider the events \GW{190412} and \GW{190521}, the first being a highly asymmetric binary~\cite{LIGOScientific:2020stg} and the second a very high-mass event~\cite{LIGOScientific:2020iuh, LIGOScientific:2020ufj}.
With these choices, our dataset contains $50$ events, which are listed in Tab.~\ref{Tab:events} in App.~\ref{sec:dataset}.
The impact of \GW{190521} on the results and other additional information about the dataset can be found in App.~\ref{App:O3_additional_results} and App.~\ref{sec:dataset}.
%
In the rest of the analysis, we adopt a \model{Madau-Dickinson} function to parametrize the rate evolution $\mathcal{R}(z)$ and a (truncated) \model{Gaussian} distribution for the mass ratio $q$.
The motivation for this choice is discussed in App.~\ref{sec:mass_ratio_rate_evolution} and summarised in Fig.~\ref{fig:violin_m2-q} therein.

\begin{figure*}
    \hspace*{-0.1cm}\includegraphics[scale=0.923]{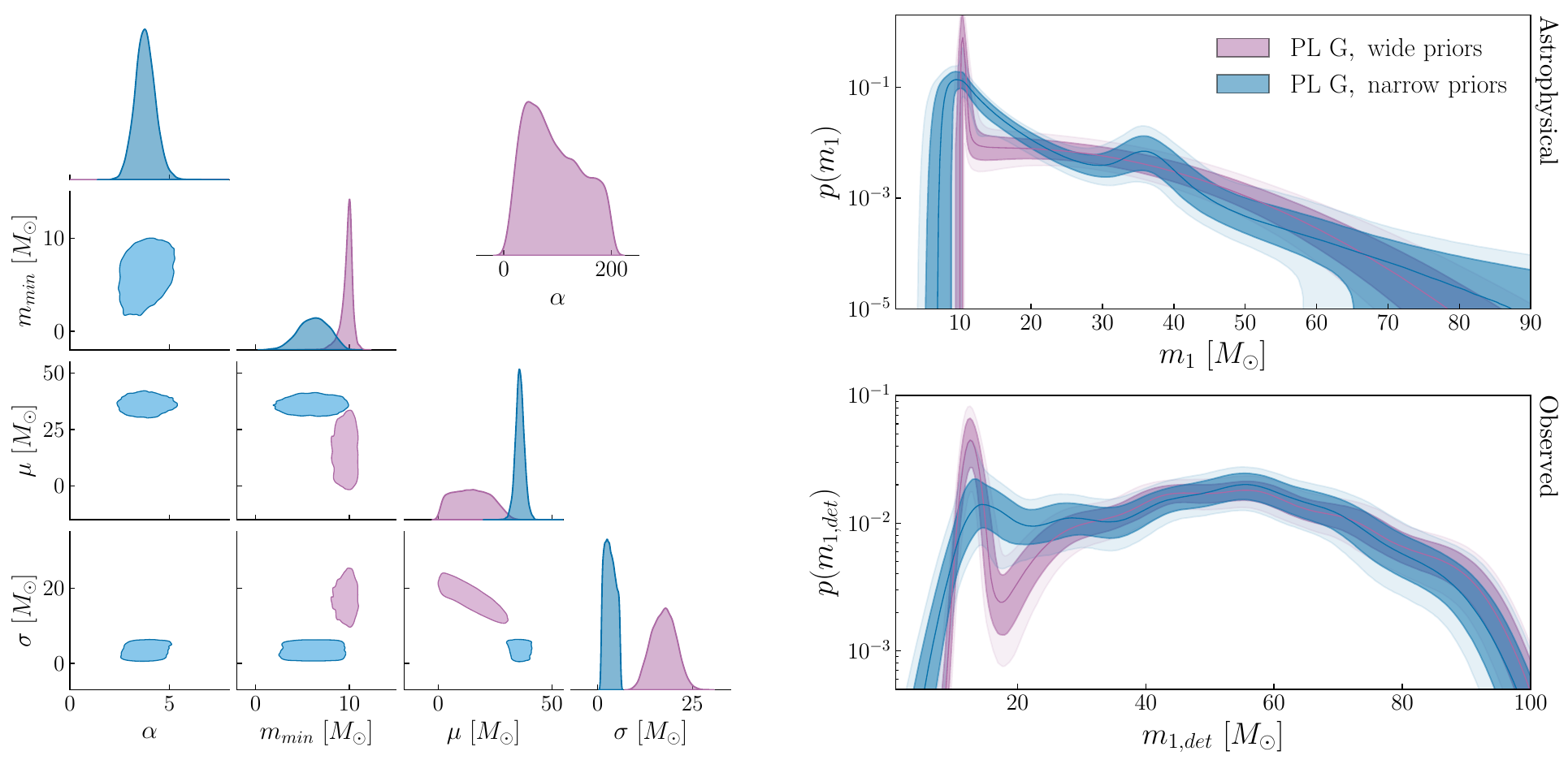}
    \caption{\justifying
    Comparison between different priors on the Powerlaw slope $\alpha$ and the Gaussian width $\sigma$ with the fiducial \model{Powerlaw-Gaussian} \footnotesize{model on O3 data. On the \textit{left}, a corner plot of these parameters together with the Powerlaw minimum $m_{min}$ and the Gaussian mean $\mu$. For illustration purposes, the posteriors on $\alpha$ with wide priors are shown separately. On the \textit{right}, the median, $68\%, 90\%$ C.I. of the reconstructed astrophysical (\textit{top panel}) and observed (\textit{bottom panel}) primary mass distribution.}}
    \label{fig:O3_stationary}
\end{figure*}

\subsection{Stationary models}
\label{Sec:stationary_models}

We first analyse O3 data with stationary models.
As extensively studied in the literature, the current fiducial model for the primary distribution from parametric analyses is a \model{Powerlaw-Gaussian}, with a mildly inclined Powerlaw peaking at $m_{min}\sim 8 M_{\odot}$ and a small Gaussian peak
at $\mu_{g}\sim 35 M_{\odot}$ ~\cite{LIGOScientific:2020kqk, KAGRA:2021duu}.
This picture, and particularly the presence of an accumulation of events around $\sim 35 M_{\odot}$, has also been confirmed by several semi- and non-parametric analyses~\cite{Tiwari:2020otp, Tiwari:2021yvr, Toubiana:2023egi, Edelman:2021zkw, Sadiq:2021fin, Farah:2023vsc, Callister:2023tgi, KAGRA:2021duu}.
Nevertheless, in this analysis we find that a \model{Powerlaw-Gaussian} model can lead to significantly different posteriors when broader priors than those typically used in the literature are considered on a few key parameters.
Specifically, by increasing the prior bounds on the Powerlaw index $\alpha$ and the Gaussian width $\sigma$, the posteriors favour larger values of both the spectral index and the Gaussian width, that are outside the standard priors used in the parametric literature.
The results are illustrated in Fig.~\ref{fig:O3_stationary},
where we show the posteriors on some relevant subset of parameters and the corresponding reconstructed distributions, astrophysical and observed, comparing different prior bounds.
The exact priors used are reported in Tab.~\ref{Tab:models_priors} in Appendix.
When wider priors are considered, the posteriors resolve a sharp peak at $m_{min}\simeq 10 M_{\odot}$ modelled by the Powerlaw, and a wide truncated Gaussian with mean $\mu \sim 20-25 M_{\odot}$ and standard deviation $\sigma \sim 15-20 M_{\odot}$.
Qualitatively, the role of the Powerlaw and the Gaussian is interchanged compared to the narrow prior analysis, so that the Gaussian covers the bulk of the distribution and the low-mass peak is fitted by the Powerlaw.
Note that the $\sim 35 M_{\odot}$ peak is not resolved anymore.
This seems to suggest that with this ``simple'' mass model, the hierarchical likelihood is multimodal, and using a narrower prior allows us to isolate a local likelihood maximum.
Most importantly, we find that the analysis with wide priors is strongly favoured, both by BF and maximum likelihood, compared to the one with the standard priors in the literature, with a (natural) logarithm of the BF of $\text{ln}\,\mathcal{B} = 5.2 \pm 0.2$ (see Tab.~\ref{Tab:stationary}).

To validate this hypothesis, we reproduce the results on simulated data, drawing events from a primary mass distribution that has both a sharp low-mass peak and an overdensity at higher masses (see Fig.~\ref{fig:O3_stationary_injection}).
As with real data, when the population has multiple features and the recovering model has only one, we observe a similar role played by the prior bounds in properly resolving both the $\sim 35 M_{\odot}$ overdensity and the $\sim 10 M_{\odot}$ peak, further indicating that the model used does not have enough features to capture those in the data.
Furthermore, in this simple simulated case, it is clear that an intermediate choice of prior bounds (\textit{medium priors} in Fig.~\ref{fig:O3_stationary_injection}) allows us to resolve two equally relevant modes in the likelihood.
Even though real data are necessarily more complex than our simulation, we believe that this simple study is able to capture the main properties associated with this model systematics.
All the details on this simulated analysis can be found in App.~\ref{App:simulations}.

\begin{center}
    \begin{tabular}{p{5cm}p{1.9cm}p{1.2cm}}
        \hline
        stationary model & $\text{ln}\,\mathcal{B}$ & $\text{ln}\,\mathcal{L}_{\text{max}}$ \\
        \hline
        &&\\[-3mm]
        \model{Powerlaw-Gaussian} \small{wide priors} & $0$ & $-1277.3$ \\
        \model{Powerlaw-Gaussian} \small{narrow priors} & $-5.2 \pm 0.2$ & $-1280.3$ \\[2mm]
        \model{Powerlaw-Gaussian-Gaussian} & $-1.1 \pm 0.2$ & $-1276.2$ \\
        \model{Powerlaw-Powerlaw-Gaussian} & $-1.1 \pm 0.2$ & $-1270.9$ \\
        \model{Powerlaw-Powerlaw-Powerlaw} & $0.3 \pm 0.2$ & $-1270.9$ \\[0.5mm]
        \hline
    \end{tabular}
    \captionof{table}{\footnotesize Table with the natural logarithm of the Bayes factor of different stationary models vs the \model{Powerlaw-Gaussian} \footnotesize model with wide priors, and the natural logarithm of the likelihood maximum.}
    \label{Tab:stationary}
\end{center}

\begin{figure}
    \hspace*{-0.15cm}\includegraphics[scale=0.51]{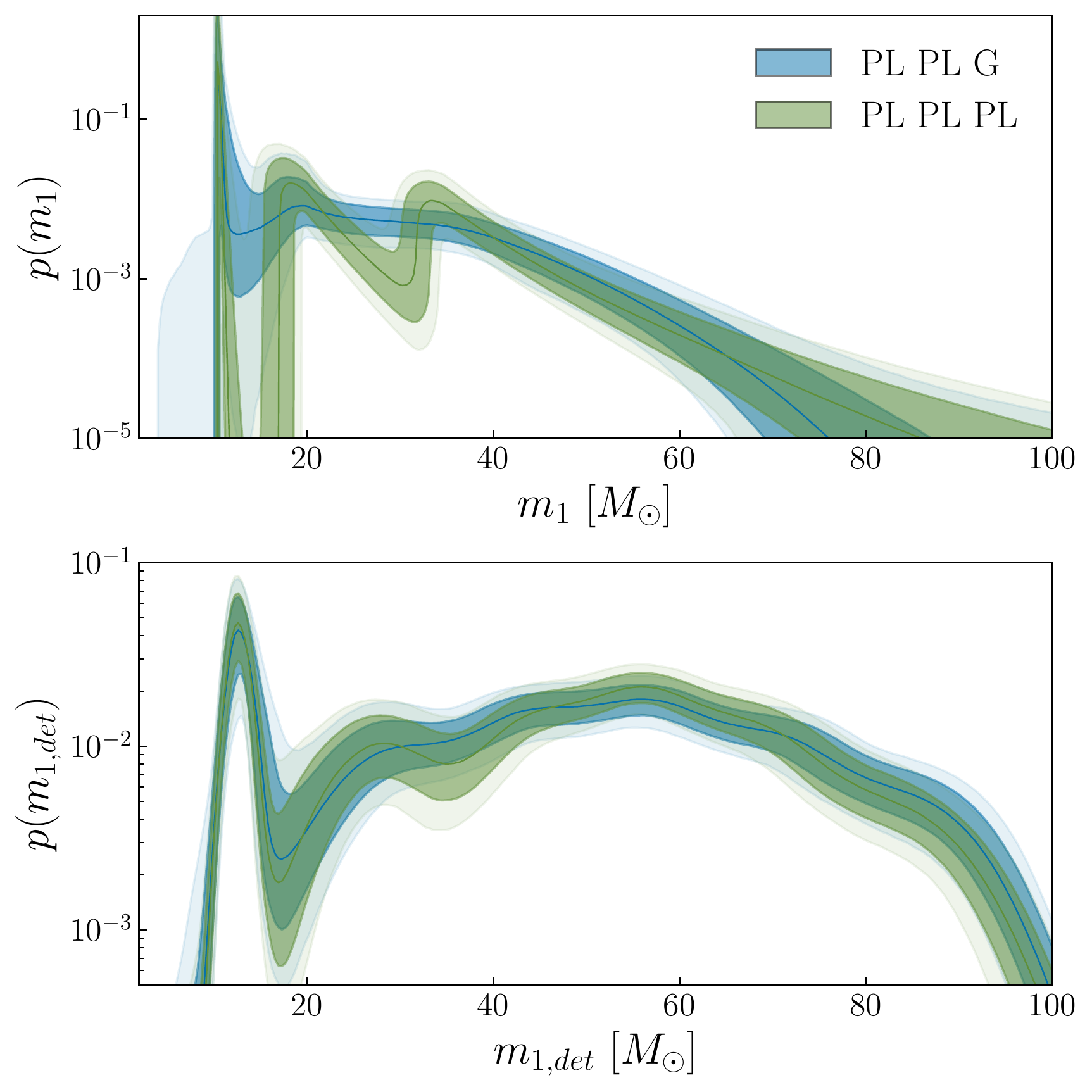}
    \caption{\justifying \footnotesize{Comparison on O3 data of the astrophysical (\textit{top panel}) and observed (\textit{bottom panel}) reconstructed primary distributions between the} \model{Powerlaw-Powerlaw-Gaussian} \footnotesize{(blue) and} \model{Powerlaw-Powerlaw-Powerlaw} \footnotesize{(green) models}. The countours mark the the median, $68\%, 90\%$ C.I.}
    \label{fig:O3_stationary_3-features}
\end{figure}

To overcome the discrepancy discussed above, as well as to resolve the $\sim 35 M_{\odot}$ peak again, we explore natural extensions of the fiducial model that include additional features.
Although we have also investigated Broken-Powerlaw models and models with only Gaussians, we only report the results of the following models: \model{Powerlaw-Gaussian-Gaussian}, \model{Powerlaw-Powerlaw-Gaussian}, \model{Powerlaw-Powerlaw-Powerlaw}.
Further details on these models, as the priors used in the analysis can be found in App.~\ref{app:models_priors}.
Fig.~\ref{fig:O3_stationary_3-features} shows the reconstructed primary distribution for the \model{Powerlaw-Powerlaw-Gaussian} and \model{Powerlaw-Powerlaw-Powerlaw} models, from which we can see that the models are indeed able to simultaneously capture the known low- and high-mass features at $\sim 10 M_{\odot}$ and $\sim 35 M_{\odot}$.
The \model{Powerlaw-Gaussian-Gaussian} model reconstructs a mass spectrum very similar to that given by the \model{Powerlaw-Gaussian}.
These results are obtained with ``wide'' priors, having ensured that they do not depend on the prior bounds (see the discussion in App.~\ref{app:models_priors}).
We report the BFs and maximum likelihood values in Tab.~\ref{Tab:stationary}, comparing the different stationary analyses with two and three features.
Only the \model{Powerlaw-Powerlaw-Powerlaw} model provides comparable evidence to the (wide priors) \model{Powerlaw-Gaussian}, with a BF of $\text{ln}\,\mathcal{B}=0.3\pm 0.2$.
As these three-feature models have more population parameters, they are naturally penalized in evidence by Occam's factor~\cite{Jefferys:1992, Jaynes:2003}, suggesting that there is not enough information in current data to distinguish them robustly.
As expected by adding more parameters, the maximum likelihood in general increases for the models with three features.

The models with three features also seem to resolve an additional peak around $\sim 20 M_{\odot}$, highlighted by a consistent drop in the population support at $\sim 15 M_{\odot}$.
This is in agreement with previous semi- and non-parametric analyses (e.g. see Fig. 11 of~\cite{LIGOScientific:2021djp}, Figs. 3-4 of~\cite{Toubiana:2023egi}, Fig. 1 of~\cite{Farah:2023vsc}, Fig. 2 of~\cite{Tiwari:2021yvr}), although the statistical robustness of this feature is still under debate in the literature~\cite{Toubiana:2023egi, Adamcewicz:2024jkr}.
Regarding our agnostic phenomenological study, we note that the presence of such sharp features in the astrophysical distribution could be partly highlighted by limitations in our parameterization.
For example, the complete absence of support for events below $10 M_{\odot}$ is probably due to the lack of flexibility in our models, which cannot extend to this region once the sharp low-mass peak is resolved, and similarly for the dip that follows.
Still, we note that the $\sim 15 M_{\odot}$ drop is consistent with the non-parametric analysis in~\cite{Toubiana:2023egi}, where the primary distribution is modelled as a collection of piece-wise Powerlaws (see the PWP model in their Figs. 3-4).

We obtain the observed distributions from the astrophysical ones by ``projecting'' onto the detectors, with a procedure that deconvolves for selection effects using a Gaussian Mixture Model (GMM).
We find that the GMM performs better in reconstructing sharp features compared to a Kernel Density Estimator (KDE), but at the cost of having ``wavy'' reconstructions as in Fig.~\ref{fig:O3_stationary} and~\ref{fig:O3_stationary_3-features}.
In general, the observed distributions seem to mitigate the differences between models in the source frame.
This behaviour is partly due to the re-weighting of the features induced by the detector sensitivity, which can only observe low-mass systems at relatively low redshifts (see Fig.~\ref{fig:O3_events_scatter}), but also to the smoothing by the GMM.

\subsection{Evolving models}
\label{Sec:O3_evolving_models}

\begin{figure*}
    \hspace*{-0.18cm}\includegraphics[scale=1.13]{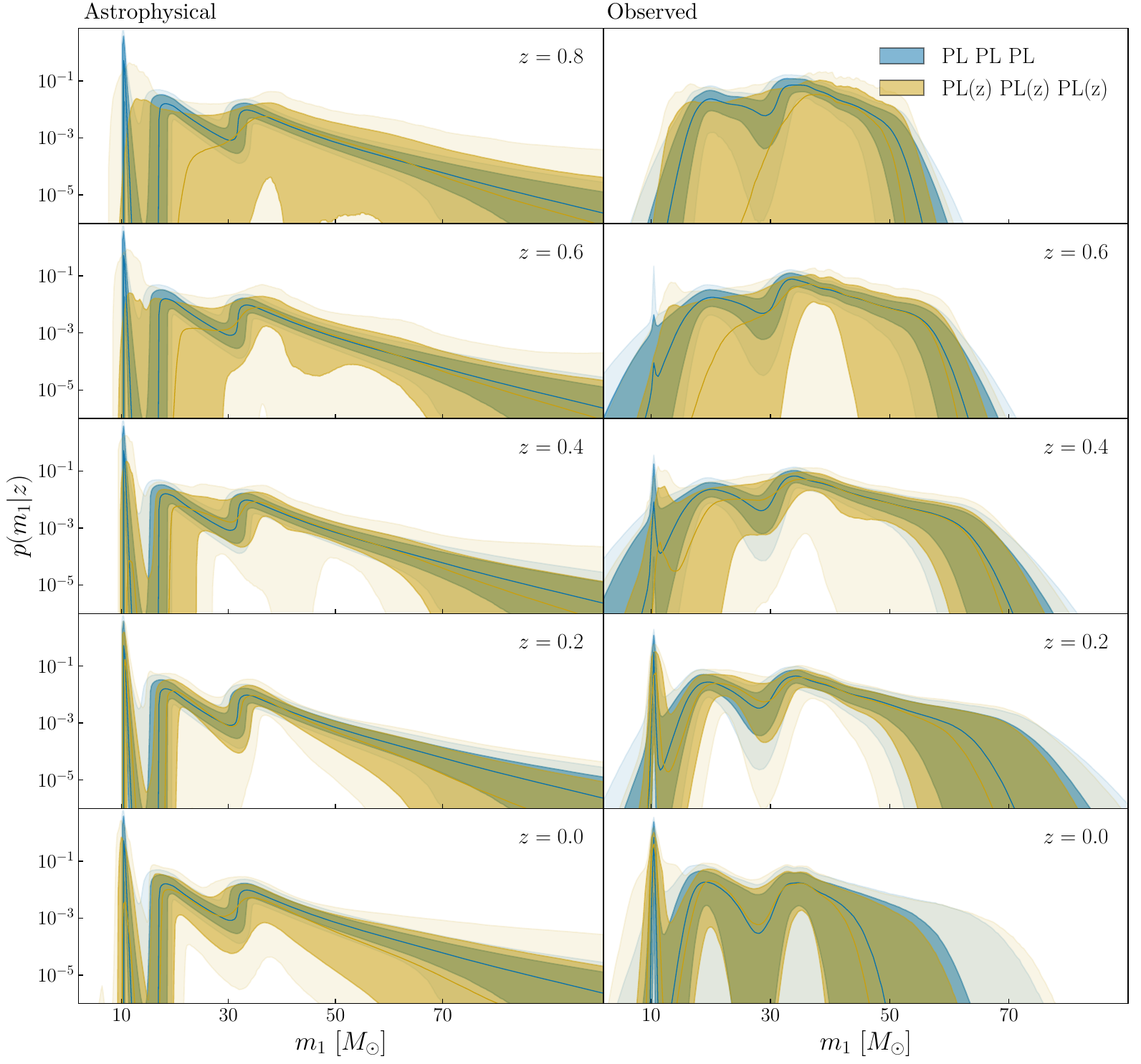}
    \caption{\justifying \footnotesize{Comparison on O3 data of the astrophysical (\textit{left panel}) and observed (\textit{right panel}) reconstructed primary distributions as a function of redshift between the stationary (blue) and redshift evolving (yellow)} \model{Powerlaw-Powerlaw-Powerlaw} \footnotesize{model}. For the evolving model, all three features evolve linearly in redshift with the two mixture functions assumed stationary. The countours mark the the median, $68\%, 90\%$ C.I.}
    \label{fig:O3_evolving}
\end{figure*}

Following the prescription in Sec.~\ref{Sec:evolving_model}, we now consider an independent linear redshift evolution of different features.
We first focus on the case where the redshift dependence in the primary mass is contained only in the features, and the mixture function is assumed to be stationary for the moment.
Results are summarised in Tab.~\ref{Tab:evolving}, where we report the maximum likelihood and the BFs of different evolving models over the two preferred stationary ones, namely the (wide priors) \model{Powerlaw-Gaussian} and the \model{Powerlaw-Powerlaw-Powerlaw}.
We do not find any significant support for the primary mass distribution to evolve with redshift.
Fig.~\ref{fig:O3_evolving_3-features} shows the posterior distributions of all the evolving parameters of the \model{Powerlaw-Powerlaw-Powerlaw} model, which are either uninformative or peaked at zero, i.e. consistent with stationarity, thus providing valuable observational constraints on the evolution of such features.
For instance, the low-mass peak is constrained to be stationary at $90\%$ C.I. as $m_{min_{a_{z_1}}} = 1.8^{+ 4.0}_{- 4.2}\, M_{\odot}$, as it is the $\sim 35 M_{\odot}$ overdensity, being $m_{min_{c_{z_1}}} = 0.7^{+ 7.5}_{- 7.6}\, M_{\odot}$ (recall that these are the linear coefficients in Eq.~\ref{eq:linear_expansion}).
We note that the redshift evolution of the third Powerlaw slope is completely unconstrained, likely reflecting the increasing missing information at higher redshifts, especially for the high-mass part of the spectrum.
Allowing the third Powerlaw maximum mass to evolve, we find largely uninformative posteriors ($m_{max_{c_{z_1}}} = 31.7^{+113.2}_{-133.1}\, M_{\odot}$, see discussion in App.~\ref{App:O3_additional_results}).

For the \model{Powerlaw-Gaussian} model, even if the evolving models are still largely disfavoured by BF, we find posteriors excluding stationarity at $90\%$ C.I. for the distribution bulk, represented by the broad truncated Gaussian (see Fig.~\ref{fig:O3_evolving_2-features}).
This behaviour seems to be in disagreement with the \model{Powerlaw-Powerlaw-Powerlaw} results and other studies in the literature,~\cite{Fishbach:2021yvy, Rinaldi:2023bbd}, which suggest a distribution that broadens at higher redshifts.
However, it is clear from the discussion in Sec.~\ref{Sec:stationary_models} that the \model{Powerlaw-Gaussian} model is unable to capture at least some of the features in the data, making difficult to understand the interplay between the evolving parameters and the biases induced by the model (similar to what was concluded in \cite{Fishbach:2021yvy}, albeit with a different dataset and model).
We also mention that, using the evolving \model{Powerlaw-Gaussian} with the narrow priors of Sec.~\ref{Sec:stationary_models}, we are not able to resolve the $\sim 35 M_{\odot}$ overdensity anymore, contrary to the stationary case.
Nonetheless, it was recently shown in~\cite{Lalleman:2025xcs} that such analysis is also possible and it still supports stationarity, assuming sufficiently narrower priors on the Powerlaw slope.

Looking at the evolving reconstructed distributions in Figs.~\ref{fig:O3_evolving} and~\ref{fig:O3_evolving_PL-G-PL3-transition}, we observe that the low-mass peak disappears at high redshifts, suggesting substantial changes in the distribution compared to the stationary case.
Although this behaviour seems to contradict the posteriors favouring stationarity in Fig.~\ref{fig:O3_evolving_3-features}, this is actually not the case.
The uncertainty in the reconstruction of the mass spectrum at high redshifts, especially for the low-mass peak at $\sim 10 M_{\odot}$, is in fact a direct consequence of the uncertainty in the posteriors of the evolving population parameters.
This effect is in fact enhanced when also the mixture function evolves with redshift, as displayed in Fig.~\ref{fig:O3_evolving_PL-G-PL3-transition}.
In this case, the posteriors on the mixture coefficients at high redshift are consistent with a weak decrease in the Peak weight, as shown in the bottom-left plot of Fig.~\ref{fig:O3_PL3_corners}.
Our inability to constrain the mass spectrum at high redshift is the product of two main limiting factors.
\begin{itemize}
    \item The low-mass, high-redshift part of the astrophysical distribution is outside the detector horizon (Fig.~\ref{fig:O3_events_scatter}).
    As mentioned in Sec.~\ref{Sec:stationary_models} for the stationary case, even if selection effects allow us to extrapolate the distribution in the region outside the detectors sensitivity, the data are only directly compared in the likelihood with the \textit{observed} distribution, which we find to be always consistent between the stationary and evolving runs (see the right panel of Figs.~\ref{fig:O3_evolving} and~\ref{fig:O3_evolving_PL-G-PL3-transition}).
    In other words, the mass spectrum reconstructed from the evolving analysis is to some extent insensitive to the presence of the low-mass peak at high redshifts.
    By contrast, for stationary models, the low-mass part of the mass spectrum reconstructed at high redshifts is only an extrapolation from the constraint that can be obtained at low redshifts.
    \item The higher the redshift, the less informative are the GW data, as the number of detected events decreases.
    This generally leads to a larger uncertainty in constraining and reconstructing the distribution for $z>0.6$, as shown in Fig.~\ref{fig:O3_evolving}.
\end{itemize}

To conclude, we briefly mention other tests and investigations that have been considered to account for possible more complex types of evolution; we report only a fraction of them for the sake of clarity.
Similar to the \model{Powerlaw-Powerlaw-Powerlaw} analysis, we included redshift evolution in the \model{Powerlaw-Gaussian-Gaussian}, \model{Powerlaw-Powerlaw-Gaussian}, \model{BrokenPowerlaw-Gaussian} models, which gave results consistent with the stationary case.
We also examined, instead of a linear, a quadratic and a powerlaw-like evolution in redshift of the Gaussian within the \model{Powerlaw-Gaussian} model, to understand whether is was possible to resolve the $\sim 35 M_{\odot}$ peak with a redshift evolving model, in case that overdensity was only present at low-redshifts.
In agreement with~\cite{Lalleman:2025xcs}, we find no support for such scenario.
Finally, we mention that the linear (or sigmoid-like) evolving mixture used cannot account for an interesting possible scenario in which a feature is present only for a certain time interval, i.e. it ``appears'' \textit{and} ``disappears'' with redshift.

\begin{center}
    \begin{tabular}{p{5cm}p{1.9cm}p{1.2cm}}
        \hline
        evolving & $\text{ln}\,\mathcal{B}$ & $\text{ln}\,\mathcal{L}_{\text{max}}$ \\
        \hline
        &&\\[-3mm]
        \model{PL-G} \small{stationary} & $0$ & $-1276.4$ \\
        \model{PL-G(z)} & $-3.1 \pm 0.2$ & $-1276.3$ \\
        \model{PL(z)-G} & $-1.9 \pm 0.2$ & $-1275.2$ \\
        \model{PL(z)-G(z)} & $-5.1 \pm 0.2$ & $-1275.3$ \\
        &&\\[-3mm]
        \hline
        &&\\[-3mm]
        \model{PL-PL-PL} \small{stationary} & $0$ & $-1270.9$ \\
        \model{PL-PL-PL(z)} & $-1.2 \pm 0.2$ & $-1272.5$ \\
        \model{PL-PL(z)-PL} & $-2.7 \pm 0.2$ & $-1273.2$ \\
        \model{PL(z)-PL-PL} & $-1.9 \pm 0.2$ & $-1272.3$ \\
        \model{PL(z)-PL(z)-PL(z)} & $-5.6 \pm 0.2$ & $-1271.5$ \\
        &&\\[-3mm]
        \hline
    \end{tabular}
    \captionof{table}{\footnotesize Table with the natural logarithm of the Bayes factor of different evolving features vs stationary analysis for the \model{Powerlaw-Powerlaw-Powerlaw} \footnotesize model, and the natural logarithm of the likelihood maximum.}
    \label{Tab:evolving}
\end{center}

\section{Discussion}\label{sec:discussion}

The results obtained in Sec.~\ref{sec:O3_data_analysis} allow us to draw some interesting conclusion on the methodology used and its potential limitations.

Bayesian model selection fundamentally relies on analysing the data under different assumptions, to statistically quantify which assumption is needed and how much it is preferred.
This is particularly significant for hierarchical parametric analyses, because the targeted physical population distribution that we aim to reconstruct is not known a priori\footnote{This differs from single-event parameter estimation, where the entire signal is in principle \textit{exactly} described by a relatively small set of parameters, and the waveform modelling is either derived directly from theory or informed by numerical simulations.}.
Even if astrophysical simulations can motivate or even predict the choice of specific population models~\cite{Talbot:2018cva, Woosley:2014lua, Vaccaro:2023cwr, Perigois:2023ihi, Torniamenti:2024uxl, Mukherjee:2021rtw}, the question of what information is actually present in the data, and to what extent we are able to extract it, still remains from a purely observational perspective.
For example, we found that the reconstruction of the low-mass part of the mass spectrum is largely uncertain for redshifts $z>0.5$.
All these issues raise the question of whether parametric analyses are a reliable approach for the future, and what population features are data-driven.
Different techniques, such as semi- or non-parametric~\cite{Mandel:2016prl, Tiwari:2020vym, Tiwari:2020otp, Edelman:2021zkw, Tiwari:2021yvr, Sadiq:2021fin, Ruhe:2022ddi, Toubiana:2023egi, Ray:2023upk, Farah:2023vsc, Callister:2023tgi, Heinzel:2023hlb, Heinzel:2024jlc, Heinzel:2024hva, Rinaldi:2023bbd}, should be able to mitigate or hopefully solve these problems, but at the cost of being more difficult to interpret physically or posing challenges in moving from the observed space to the astrophysical space~\cite{Ng:2024xps, Fabbri:2025faf, Farah:2024xub}.

Although different approaches are possible~\cite{LIGOScientific:2017adf, LIGOScientific:2019zcs, LIGOScientific:2021aug}, most of the information from current cosmological analyses with GWs without counterparts comes from the BH mass distribution~\cite{Mastrogiovanni:2022hil, Mastrogiovanni:2023emh, Mastrogiovanni:2023zbw, Gray:2019ksv, Gray:2021sew, Gray:2023wgj}, whose features allow to extract information about the large-scale dynamics of the Universe, most notably the \textit{Hubble parameter}.
A faithful characterization of the mass distribution is therefore essential to obtain unbiased measurements of cosmological parameters, especially with respect to the mismodelling of the redshift evolution~\cite{Pierra:2023deu, Agarwal:2024hld, Roy:2024oxh}.
Although in this study we find no evidence for redshift evolution in the current data, the effect of correlations between intrinsically evolving features in the astrophysical distribution and cosmology remains to be investigated, since both effects give rise to qualitatively similar behaviour in the observed distribution.
In fact, if time evolution is negligible at current sensitivities, future detectors will observe BH mergers up to very high redshifts~\cite{LISA:2024hlh, Tamanini:2016zlh, Laghi:2021pqk, Muttoni:2021veo}, making the modelling of redshift evolution an essential concern.
Such analysis is beyond the scope of this work, and will be addressed in future studies.

\section{Conclusions}\label{sec:conclusions}

We presented an extensive parametric analysis on the population of the GW events observed in the third LVK observing run, O3, using three-features, redshift-dependent mass models for the primary BH, which allow to phenomenologically capture additional structure and possible evolution of the population on cosmological times.
Through Bayesian model selection, we systematically analyse the data under different model assumptions, providing an exhaustive characterization of i) the feature content of the primary BH distribution, ii) the redshift evolution of such features.
In the stationary case, we find that the results with the fiducial \model{Powerlaw-Gaussian} depend on the prior bounds used, with the wide prior analysis largely favoured over the narrower priors usually found in the literature, by $\text{ln}\,\mathcal{B}=5.2 \pm 0.2$.
This effect is caused by the presence of multiple features in the data that cannot be captured simultaneously with a \model{Powerlaw-Gaussian} model, as reproduced with a simulated population.
We find that models with three features are able to resolve multiple features, in agreement with existing semi- and non-parametric results in the literature, with the \model{Powerlaw-Powerlaw-Powerlaw} model being even slightly preferred over the wide priors $\text{ln}\,\mathcal{B}=0.3 \pm 0.2$; to our knowledge, this is the first time that such a parametrization has been used.

We find no evidence for redshift evolution with any parametrization adopted, and provide constraints on the possible linear redshift evolution of the \model{Powerlaw-Powerlaw-Powerlaw} model features.
The evolving \model{Powerlaw-Gaussian} model shows weak support for changes in the bulk of the distribution at higher redshifts.
We highlight a peculiar behaviour of the evolving features outside the detector horizon: since low-mass and high-redshift events are not detected, the results are to some extent insensitive to changes in that part of the distribution.
This points to inherent limitations of hierarchical parametric analyses in reconstructing astrophysical distributions beyond the detector horizon with time-dependent models.
Future analyses with upcoming GW observations will be able to assess the statistical robustness of our results.

Our method could be further improved by including the redshift dependence of the secondary mass, correlations between the primary mass and the mass ratio, and by accounting for the BH spins.
Evidence for population-level correlations of the spins with the mass ratio and primary mass has already been observed, raising the question of the robustness of such results when the redshift evolution of these parameters is taken into account.
Such studies could help to shed light on the astrophysical formation channels of the observed BH mergers, however these steps will come with significant challenges in identifying appropriate parametrizations, interpretation of results, and computational cost.
These difficulties could be mitigated by using other, more agnostic approaches, such as semi- or non-parametric methods, which may limit the systematics derived from the model assumptions.

Our work has important implications for cosmological analyses with GWs, which fundamentally rely on a faithful characterization of the population structure to simultaneously infer the cosmological parameters.
Consequently, these studies are particularly sensitive to analysis bias, especially if the model does not properly account for redshift evolution in the data~\cite{Pierra:2023deu, Agarwal:2024hld}.
If on the one hand our results do not support the redshift evolution of the primary BH mass with current events, they also point to systematics that may affect future cosmological analyses with time-evolving population distributions.
These questions will be addressed in future studies.

Our work serves as a solid foundation for hierarchical population analyses with time-dependent models, both in the ongoing LVK observing run (O4) and with future more sensitive GW detectors, nearing the possibility of
astrophysics and cosmology with more realistic phenomenological black hole distributions.

\begin{acknowledgments}

We thank Jack Heinzel for his valuable comments on the manuscript, and Alexandre Toubiana, Stefano Rinaldi, Walter Del Pozzo for useful discussions on GWTC-3 non-parametric results.
V.G. would like to thank Ollie Burke, Manuel Piarulli and Gergely Dálya for fruitful conversations throughout the project.
This paper was reviewed by the LIGO-Virgo-KAGRA Collaboration (document number: P2500059).
%
V.G, N.T. and S.Marsat acknowledge support form the French space agency CNES in the framework of LISA.
This project has received financial support from the CNRS through the AMORCE funding framework and from the Agence Nationale de la Recherche (ANR) through the MRSEI project ANR-24-MRS1-0009-01.
S. Mastrogiovanni is supported by ERC Starting Grant No. 101163912–GravitySirens.
%
This research made use of data, software and/or web tools obtained from the Gravitational Wave Open Science Center~\cite{LIGOScientific:2019lzm, KAGRA:2023pio}, a service of the LIGO Scientific Collaboration, the KAGRA Collaboration and the Virgo Collaboration.
%
The authors are grateful for computational resources provided by the LIGO Laboratory (LHO) and supported by National Science Foundation Grants PHY-0757058 and PHY-0823459. This material is based upon work supported by NSF's LIGO Laboratory which is a major facility fully funded by the National Science Foundation.
%
LIGO Laboratory and Advanced LIGO are funded by the United States National Science Foundation (NSF) as well as the Science and Technology Facilities Council (STFC) of the United Kingdom, the Max-Planck Society (MPS), and the State of Niedersachsen/Germany for support of the construction of Advanced LIGO and construction and operation of the GEO600 detector. Additional support for Advanced LIGO was provided by the Australian Research Council. Virgo is funded, through the European Gravitational Observatory (EGO), by the French Centre National de Recherche Scientifique (CNRS), the Italian Istituto Nazionale di Fisica Nucleare (INFN) and the Dutch Nikhef, with contributions by institutions from Belgium, Germany, Greece, Hungary, Ireland, Japan, Monaco, Poland, Portugal, Spain. KAGRA is supported by Ministry of Education, Culture, Sports, Science and Technology (MEXT), Japan Society for the Promotion of Science (JSPS) in Japan; National Research Foundation (NRF) and Ministry of Science and ICT (MSIT) in Korea; Academia Sinica (AS) and National Science and Technology Council (NSTC) in Taiwan.\\

\noindent{\textbf{\textit{Software}}.} The plots have been produced using the open source \texttt{python} package \texttt{namib}~\cite{namib}, which makes use of the \texttt{statsmodels}~\cite{seabold2010statsmodels} and \texttt{joypy}~\cite{joypy} packages.
The \texttt{ICAROGW}~\cite{icarogw} inference relies on the publicly available packages: \texttt{bilby}~\cite{Ashton:2018jfp, Romero-Shaw:2020owr, Smith:2019ucc} and the \texttt{dynesty}~\cite{Speagle:2019ivv, dynesty, Skilling2006NestedSF} sampler.
The manuscript content has been derived using publicly available software: \texttt{matplotlib, numpy, scipy, seaborn}~\cite{matplotlib, numpy, scipy, seaborn}.
\end{acknowledgments}

\newpage
\appendix
\renewcommand{\thefigure}{A.\arabic{figure}} 
\setcounter{figure}{0}  
\section*{Appendix}

All the results are obtained assuming a standard $\Lambda$CDM cosmology and fixing the cosmological parameters to those obtained by the Planck Collaboration~\cite{Planck:2015fie}, i.e. $H_0 = 67.7$ and $\Omega_{m} = 0.308$.

\section{Models and priors}\label{app:models_priors}

In this section, we discuss the functional forms used for the different models and their priors.\\

\textit{Primary mass --}
The underlying functional forms of the models used are described in~\cite{Mastrogiovanni:2023zbw}.
The Powerlaw feature is defined in App. B.2.1 (Truncated Power-Law), the Gaussian in App. B.2.
The Gaussian is truncated at the low-mass bound of the Powerlaw, $m_{min}$.
In our framework, these basic blocks are combined to form multiple feature models through mixture coefficients (e.g. in Eq. B.15 of~\cite{Mastrogiovanni:2023zbw}, which corresponds to our Eq.~\eqref{evolving_primary_mass}).
When more than two features are present, additional coefficients are introduced, such that the total probability is still normalised, $\lambda_1 + ... + \lambda_n = 1$.
For example, a model with three features has two mixture coefficients, $\lambda_{\alpha}$ and $\lambda_{\beta}$, such that the weight of the third feature is $1-\lambda_{\alpha}-\lambda_{\beta}=0$.
For proper normalization, we ensure that these coefficients are bounded between $[0,1]$, and that $\lambda_{\alpha}+\lambda_{\beta}<1$.
To include redshift evolution, we redshift-expand the population parameters as in Eq.~\eqref{eq:linear_expansion}.
The resulting distribution is \textit{conditioned} on redshift, meaning that for each $z$, $p_{\text{pop}}(m_1\,|\,z,\vb{\Lambda})$ is a normalized density distribution.
We use a different parametrization for the mixture coefficients (see Sec.~\ref{Sec:evolving_model}), which we parametrize as a line in a $[0,1]\times[0,1]$ box,
\begin{equation}
    \text{mix}(z) = (\text{mix}_{z_1} - \text{mix}_{z_0})z + \text{mix}_{z_0},
\end{equation}
having defined $\text{mix}_{z_0}\equiv \text{mix}(0)$ and $\text{mix}_{z_1}\equiv \text{mix}(1)$.\\

\textit{Secondary mass and rate evolution --}
We parametrize the secondary mass through the mass ratio, $q\equiv m_2/m_1$, for which we adopt a truncated \model{Gaussian} (App. B.2 of~\cite{Mastrogiovanni:2023zbw}).
As described in Sec.~\ref{Sec:evolving_factorisation}, we assume the mass ratio to be uncorrelated with both the primary mass and the redshift, so that $p_{\text{pop}}(q\,|\,m_1,z,\vb{\Lambda}) = p_{\text{pop}}(q\,|\,\vb{\Lambda})$.
For the rate evolution we use the \model{Madau-Dickinson} function defined in Eq. B.2 of~\cite{Mastrogiovanni:2023zbw}.
Specifically, the \model{Madau-Dickinson} parametrizes the function $\psi(z)$, defined as $\mathcal{R}(z)\equiv \mathcal{R}_0\, \psi(z)$ (see Eq. 15 of~\cite{Mastrogiovanni:2023zbw}).
The motivation for using such parametrizations is described in App.~\ref{sec:mass_ratio_rate_evolution}.\\

\textit{Priors --}
The priors used for the different models can be found in Tab.~\ref{Tab:models_priors}.
We impose an ordering on the features in the primary mass spectrum by using prior bounds that prevent complete overlap of the features in mass.
Without such a constraint, the degeneracy resulting from the exchange of features is reflected in the likelihood by multimodalities.
The symmetry is exact for identical features, such as in the \model{Powerlaw-Powerlaw-Powerlaw} model.
We ensure that our choice of prior bounds does not affect the results by running all the analyses without any prior ordering.

We also consider broader priors for the rate evolution parameters than those typically found in the literature.
The resulting posteriors (see Fig.~\ref{fig:O3_mass-ratio_rate-evolution}) admit a large support for negative values of $\gamma$, reflecting the fact that for $\kappa \sim 0$ the \model{Madau-Dickinson} is effectively independent for values where $\gamma<0$.
The support for large values of $z_p$ reflects the uncertainty in constraining the rate evolution at high redshifts.
Nevertheless, we note that for both $\gamma$ and $z_p$ a peak of higher probability is clearly distinguishable in the posteriors, mostly independent of the model used.

\begin{table*}
\setlength{\tabcolsep}{8pt}
\scalebox{0.8}{
    \begin{tabular}{l l}
    
        \bf{Primary stationary}&\\
        \hline
        &\\[-1mm]
        
        \bf{\model{Powerlaw-Gaussian}} \textnormal{wide priors} & \\
        & \\[-4mm]
        \begin{tabular}{ccccccc}
            $\alpha$ & $m_{\text{min}}$ & $m_{\text{max}}$ & $\delta_m$ & $\mu$ & $\sigma$ & $\text{mix}$ \\
            $(-4, 200)$ & $(1, 100)$ & $(30, 200)$ & $(0, 10)$ & $(1, 60)$ & $(1, 30)$ & $(0, 1)$ \\
        \end{tabular} & \\[3mm]
        & \\[0mm]
        
        \bf{\model{Powerlaw-Gaussian}} \textnormal{narrow priors} & \\
        & \\[-4mm]
        \begin{tabular}{ccccccc}
            $\alpha$ & $m_{\text{min}}$ & $m_{\text{max}}$ & $\delta_m$ & $\mu$ & $\sigma$ & $\text{mix}$ \\
            $(-4, \:\:12)$ & $(1, 100)$ & $(30, 200)$ & $(0, 10)$ & $(1, 60)$ & $(1, \:\:6)$ & $(0, 1)$ \\
        \end{tabular} & \\[3mm]
        & \\[0mm]
        
        \bf{\model{Powerlaw-Gaussian-Gaussian}} & \\
        & \\[-4mm]
        \begin{tabular}{ccccccccc}
            $\alpha$ & $m_{\text{min}}$ & $m_{\text{max}}$ & $\mu_a$ & $\sigma_a$ & $\mu_b$ & $\sigma_b$ & $\text{mix}_{\alpha}$ & $\text{mix}_{\beta}$ \\
            $(-4, 200)$ & $(1, \:\:15)$ & $(30, 200)$ & $(10, 20)$ & $(1, 30)$ & $(20, 60)$ & $(1, 30)$ & $(0, 1)$ & $(0, 1)$ \\
        \end{tabular} & \\[3mm]
        & \\[0mm]
        
        \bf{\model{Powerlaw-Powerlaw-Gaussian}} & \\
        & \\[-4mm]
        \begin{tabular}{cccccccccc}
            $\alpha_a$ & $m_{\text{min}_a}$ & $m_{\text{max}_a}$ & $\alpha_b$ & $m_{\text{min}_b}$ & $m_{\text{max}_b}$ & $\mu$ & $\sigma$ & $\text{mix}_{\alpha}$ & $\text{mix}_{\beta}$ \\
            $(-4, 200)$ & $(1, \:\:15)$ & $(30, 200)$ & $(-4, 20)$ & $(10, 20)$ & $(30, 200)$ & $(20, 60)$ & $(1, 30)$& $(0, 1)$ & $(0, 1)$ \\
        \end{tabular} & \\[3mm]
        & \\[0mm]
        
        \bf{\model{Powerlaw-Powerlaw-Powerlaw}} & \\
        & \\[-4mm]
        \begin{tabular}{ccccccccccc}
            $\alpha_a$ & $m_{\text{min}_a}$ & $m_{\text{max}_a}$ & $\alpha_b$ & $m_{\text{min}_b}$ & $m_{\text{max}_b}$ & $\alpha_c$ & $m_{\text{min}_c}$ & $m_{\text{max}_c}$ & $\text{mix}_{\alpha}$ & $\text{mix}_{\beta}$ \\
            $(-4, 200)$ & $(1, \:\:15)$ & $(30, 200)$ & $(-4, 50)$ & $(10, 20)$ & $(30, 200)$ & $(-4, 50)$ & $(15, 60)$ & $(30, 200)$ & $(0, 1)$ & $(0, 1)$ \\
        \end{tabular} & \\[8mm]

        \bf{Primary evolving}&\\
        \hline
        &\\[-1mm]
        
        \bf{\model{Powerlaw-Gaussian}} & \\
        & \\[-4mm]
        \begin{tabular}{cccccc}
            $\alpha_{z_0}$ & $m_{\text{min}_{z_0}}$ & $m_{\text{max}_{z_0}}$ & $\mu_{z_0}$ & $\sigma_{z_0}$ & $\text{mix}_{z_0}$ \\
            $(-4, 200)$ & $(1, 100)$ & $(30, 200)$ & $(1, 60)$ & $(1, 30)$ & $(0, 1)$ \\
            $\alpha_{z_1}$ & $m_{\text{min}_{z_1}}$ & $m_{\text{max}_{z_1}}$ & $\mu_{z_1}$ & $\sigma_{z_1}$ & $\text{mix}_{z_1}$ \\
            $(-200, 200)$ & $(-100, 100)$ & 0 & $(-200, 200)$ & $(0, 200)$ & $(0, 1)$ \\
        \end{tabular} & \\[3mm]
        & \\[0mm]

        \bf{\model{Powerlaw-Powerlaw-Powerlaw}} & \\
        & \\[-4mm]
        \begin{tabular}{ccccccccccc}
            $\alpha_{a_{z_0}}$ & $m_{\text{min}_{a_{z_0}}}$ & $m_{\text{max}_{a_{z_0}}}$ & $\alpha_{b_{z_0}}$ & $m_{\text{min}_{b_{z_0}}}$ & $m_{\text{max}_{b_{z_0}}}$ & $\alpha_{c_{z_0}}$ & $m_{\text{min}_{c_{z_0}}}$ & $m_{\text{max}_{c_{z_0}}}$ & $\text{mix}_{\alpha_{z_0}}$ & $\text{mix}_{\beta_{z_0}}$ \\
            $(-4, 120)$ & $(1, \:\:20)$ & $(30, 200)$ & $(-4, 20)$ & $(15, 25)$ & $(30, 200)$ & $(-4, 20)$ & $(25, 60)$ & $(30, 200)$ & $(0, 1)$ & $(0, 1)$ \\
            $\alpha_{a_{z_1}}$ & $m_{\text{min}_{a_{z_1}}}$ & $m_{\text{max}_{a_{z_1}}}$ & $\alpha_{b_{z_1}}$ & $m_{\text{min}_{b_{z_1}}}$ & $m_{\text{max}_{b_{z_1}}}$ & $\alpha_{c_{z_1}}$ & $m_{\text{min}_{c_{z_1}}}$ & $m_{\text{max}_{c_{z_1}}}$ & $\text{mix}_{\alpha_{z_1}}$ & $\text{mix}_{\beta_{z_1}}$ \\
            $(-100, 100)$ & $(-100, 100)$ & $0$ & $(-100, 100)$ & $(-100, 100)$ & 0 & $(-100, 100)$ & $(-100, 100)$ & 0 & $(0, 1)$ & $(0, 1)$ \\
        \end{tabular} & \\[8mm]
        & \\[0mm]
        
        \bf{Mass ratio \& rate evolution}&\\
        \hline
        &\\[-1mm]
        
        \bf{\model{Gaussian}} \textnormal{(truncated)} & \\
        & \\[-4mm]
        \begin{tabular}{cc}
            $\mu_q$ & $\sigma_q$ \\
            $(0.1, 1)$ & $(0.01, 0.9)$ \\
        \end{tabular} & \\[3mm]
        & \\[0mm]

        \bf{\model{Madau-Dickinson}} & \\
        & \\[-4mm]
        \begin{tabular}{cccc}
            $\gamma$ & $\kappa$ & $z_p$ & $\mathcal{R}_0$ \\
            $(-50, 30)$ & $(-20, 10)$ & $(0, 4)$ & $(0, 100)$ \\
        \end{tabular} & \\[3mm]
        \hline
        
    \end{tabular}
}
\caption{\footnotesize Priors for the population parameters of the different models considered.}
\label{Tab:models_priors}
\end{table*}

\section{Mass ratio and rate evolution}\label{sec:mass_ratio_rate_evolution}

In this section we report the model selection results for the mass ratio distribution and rate evolution that we performed to select the standard models used in the O3 analysis, and show their observed values for different primary models.\\

\textit{Model selection --}
For the mass ratio, we consider a \model{Powerlaw} and a truncated \model{Gaussian} model.
Note that the truncated \model{Gaussian} still allows to fit distributions that have a maximum at $q\sim 1$, but can also capture a peak at $q<1$, unlike the \model{Powerlaw} model.
For rate evolution, we consider a \model{Powerlaw} and a \model{Madau-Dickinson} function.
The \model{Madau-Dickinson} is a phenomenological function describing the star formation rate~\cite{Madau:2014bja}, which behaves like a \model{Powerlaw} at low redshifts, but peaks at a certain redshift $z_p$.

\begin{figure}[h!]
    \hspace*{-0.5cm}\includegraphics[scale=0.5]{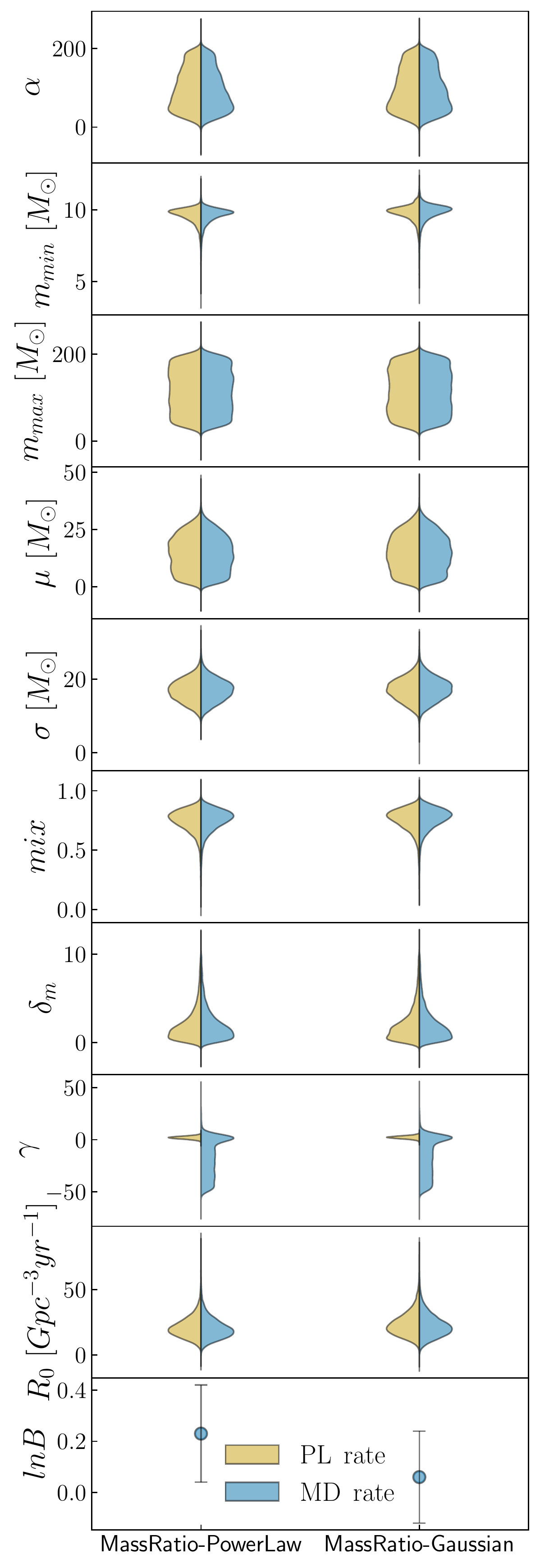}
    \caption{\justifying \footnotesize
    Violin plot comparing the posterior distributions of the population parameters for the \model{Powerlaw-Gaussian} \footnotesize primary model, using a \model{Powerlaw} \footnotesize (yellow) or a \model{Madau-Dickinson} \footnotesize function for the rate evolution $\psi(z)$ on O3 data.
    The two columns have different parametrizations for the mass ratio, namely a \model{Powerlaw} \footnotesize (\textit{left}) and a \model{Gaussian} \footnotesize (\textit{Right}).
    The last row shows the natural logarithm of the  Bayes factor for the \model{Madau-Dickinson} \footnotesize vs the \model{Powerlaw} \footnotesize hypothesis.}
    \label{fig:violin_m2-q}
\end{figure}

Assuming a (wide prior) stationary \model{Powerlaw-Gaussian} model for the primary mass, we consider all combinations of these models for the mass ratio and rate evolution, and compute BFs for model selection.
Fig.~\ref{fig:violin_m2-q} compares the posterior distributions on the common population parameters between the \model{Powerlaw} (yellow) and the \model{Madau-Dickinson} (blue), assuming a \model{Powerlaw} (left) and a \model{Gaussian} (right) for the mass ratio.
The last row shows the BFs for the different hypotheses.
Both the posteriors and the BFs show that the different parameterizations are basically indistinguishable with the O3 data.
Given this uncertainty, we decide to use the more flexible parameterizations in our study, i.e. the \model{Gaussian} model for the mass ratio and the \model{Madau-Dickinson} model for the rate evolution.
In fact, more degrees of freedom make the model more flexible for unexpected behaviours when considering different primary mass models.

\begin{figure*}
    \hspace*{-0.4cm}\includegraphics[scale=0.9]{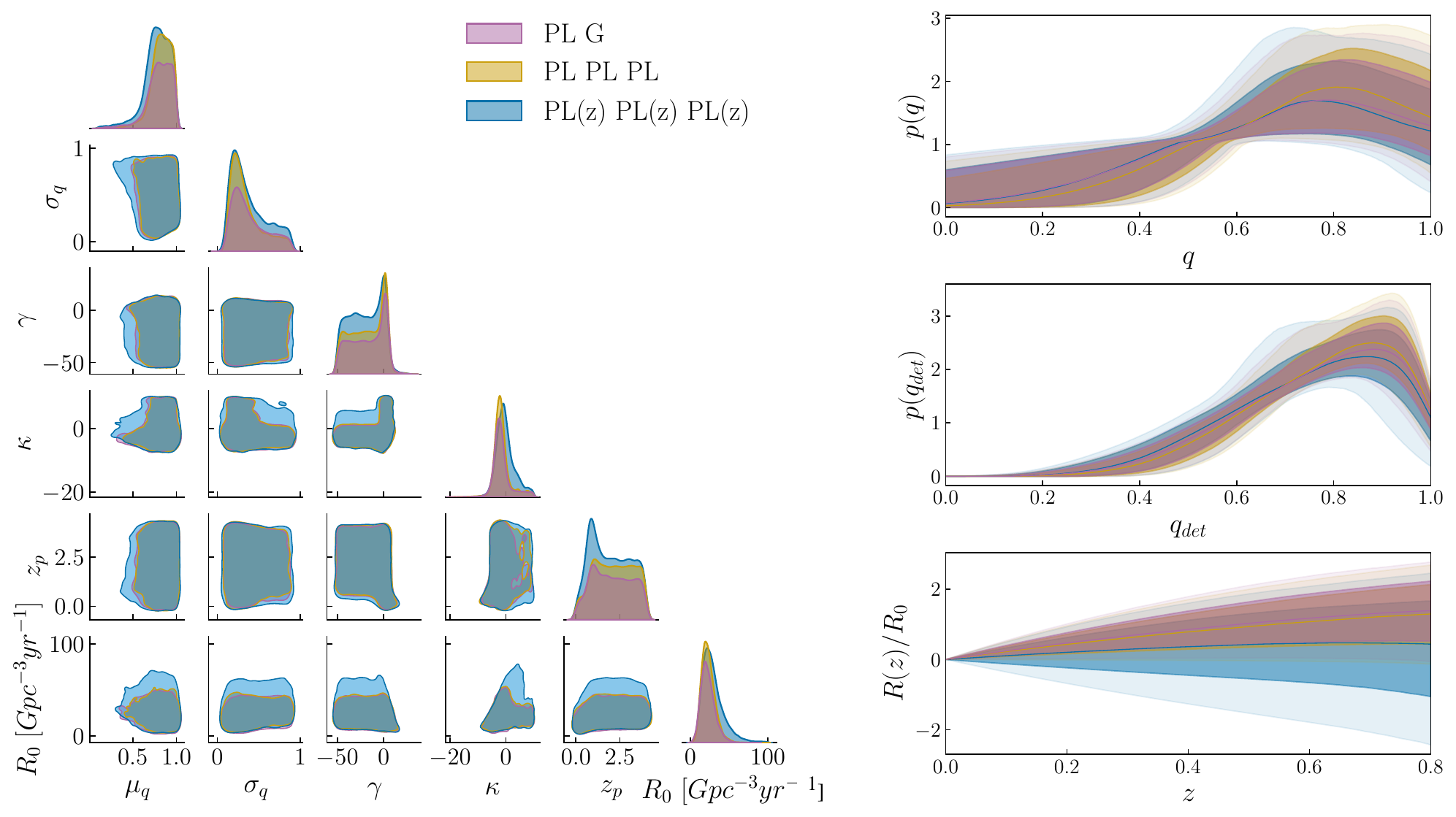}
    \caption{\justifying \textit{Left}: corner plot with the posteriors on the mass ratio and rate evolution population parameters, for different primary models on O3 data.
    \textit{Right}: the median, $68\%, 90\%$ C.I. of the reconstructed astrophysical (\textit{top}) and observed (\textit{middle}) mass ratio distributions, and astrophysical rate evolution function (\textit{bottom}).}
    \label{fig:O3_mass-ratio_rate-evolution}
\end{figure*}

Finally, we mention that in order to reconstruct the observed mass ratio spectrum in Fig.~\ref{fig:O3_mass-ratio_rate-evolution}, we have ensured that the peak at $q\sim 0.9$ does not depend on selection effects, with the detection probability being relatively constant on $q$.\\

\textit{O3 posteriors --}
In Fig.~\ref{fig:O3_mass-ratio_rate-evolution}, we show the posteriors on the mass ratio and rate evolution population parameters (left), comparing the results from different primary models, both stationary and evolving.
On the right, the reconstructed astrophysical (top) and observed (middle) mass ratio distributions, and the astrophysical rate evolution function (bottom).
The results are consistent across different primary models.
The redshift evolving \model{Powerlaw-Powerlaw-Powerlaw} recovers a slightly lower rate at higher redshifts, and appears to resolve the \model{Madau-Dickinson} redshift peak, $z_p$, better.
This is consistent with the uncertainty in reconstructing the low-mass peak at high redshift outside the detector horizon, as discussed in Sec.~\ref{Sec:O3_evolving_models}.
We also note that the posteriors at $\mu_q$ and $\sigma_q$ appear to resolve a peak at $q\sim 0.8$ better, although there is clearly not enough information in the data to draw any conclusions.

\section{Additional details on simulations}\label{App:simulations}

This section provides details of the setup and settings used for the simulated data analyses.
Following~\cite{Pierra:2023deu}, for both the selection effect injections and the simulated events, we set the detection threshold to $\text{SNR}_{\text{det}}=12$.
The SNR is computed with the procedure described in Sec. II.A of~\cite{Pierra:2023deu}.
We estimate the matched filter SNR using an inspiral-like approximation for the optimal part and adding a unit Gaussian random variable to mimic the effect of detector noise.
For more details on the motivation for using this approximation and its limitations, see~\cite{Allen:2005fk, Fishbach:2019ckx, Essick:2023toz, Essick:2023upv, Mastrogiovanni:2022hil, Pierra:2023deu}.
This approximation is sufficient for our purposes of testing the pipeline, as we are not interested in a realistic simulation campaign.
Instead of drawing posterior samples from the simulated events, we use the \textit{true values} directly for hierarchical inference.
Although this is not realistic, it allows us to remove additional uncertainty from the posteriors of the population parameters, while still being consistent with our implementation of selection effects~\cite{Essick:2023upv}.
We generate $10^5$ injections to compute selection effects, using a \model{Powerlaw-Gaussian} model that is slightly ``broader'' than the injected population to ensure proper coverage of the population parameter space.\\

\textit{Evolution validation --}
We validate the implementation of the redshift evolution using the \model{Powerlaw-Gaussian} model.
We draw events from different injected populations, both stationary and redshift evolving.
The following population parameters are fixed for all cases,
\begin{align*}
    &\alpha = 3.8, m_{\text{min}} = 7, m_{\text{max}} = 150, \delta_m = 5,\\
    &\mu_{z_0} = 35, \sigma_{z_0} = 6, \sigma_{z_1} = 0, \text{mix}_{z_0} = 0.9,\\ &\text{mix}_{z_1} = 0.9, \mu_{q} = 0.8, \sigma_{q} = 0.15,
\end{align*}
and their values are qualitatively taken from the results of the LVK GWTC-3 analysis~\cite{KAGRA:2021duu}.
Assuming a linear evolution of the Gaussian peak and a \model{Powerlaw} rate evolution, we consider the following cases of injection
\begin{enumerate}
    \item $\mu_{z_1} = 0$, $\gamma = 0$,
    \item $\mu_{z_1} = 0$, $\gamma = 3$,
    \item $\mu_{z_1} = 30$, $\gamma = 0$,
    \item $\mu_{z_1} = 30$, $\gamma = 3$,
\end{enumerate}
with $\gamma$ being the slope of the rate evolution.
For each case, we recover with a stationary (S) and an evolving (E) model for the primary mass.
The rate evolution is always left free to vary.
If most of these cases are simply consistency tests, those in which we recover evolving data with a stationary model are actually studies of systematics.
To account for the Poisson noise of drawing a finite number of events, we analyse five population realizations for each case.
To test the model at different sensitivities, we generate an ``O3-realistic'' population with an O3-like number of detected events, and an ``O3-enhanced'' population, which allows us to obtain sharp posteriors on the population parameters and highlight possible systematics within the pipeline.
For the cases above, these are the number of generated events $N_{\text{gen}}$  and the number of detected events $N_{\text{det}}$ (the latter averaged over the five population realizations)
\begin{enumerate}
    \item \begin{itemize}
        \item \small{O3-realistic: $N_{\text{gen}} = 18000$, $N_{\text{det}} \simeq 56$
        \item O3-enhanced: $N_{\text{gen}} = 350000$, $N_{\text{det}} \simeq 1057$}
    \end{itemize}
    \item \begin{itemize}
        \item \small{O3-realistic: $N_{\text{gen}} = 10000$, $N_{\text{det}} \simeq 61$
        \item O3-enhanced: $N_{\text{gen}} = 1666666$, $N_{\text{det}} \simeq 1002$}
    \end{itemize}
    \item \begin{itemize}
        \item \small{O3-realistic: $N_{\text{gen}} = 18000$, $N_{\text{det}} \simeq 60$
        \item O3-enhanced: $N_{\text{gen}} = 350000$, $N_{\text{det}} \simeq 1340$}
    \end{itemize}
    \item \begin{itemize}
        \item \small{O3-realistic: $N_{\text{gen}} = 60000$, $N_{\text{det}} \simeq 54$
        \item O3-enhanced: $N_{\text{gen}} = 1200000$, $N_{\text{det}} \simeq 1017$}
    \end{itemize}
\end{enumerate}
\normalsize{}
For reasons of space, we only report the results of the case 4., which are described in Sec.~\ref{Sec:simulations}, but we mention that all analyses produced results in agreement with expectations.\\

\begin{figure*}
    \hspace*{-0.1cm}\includegraphics[scale=0.91]{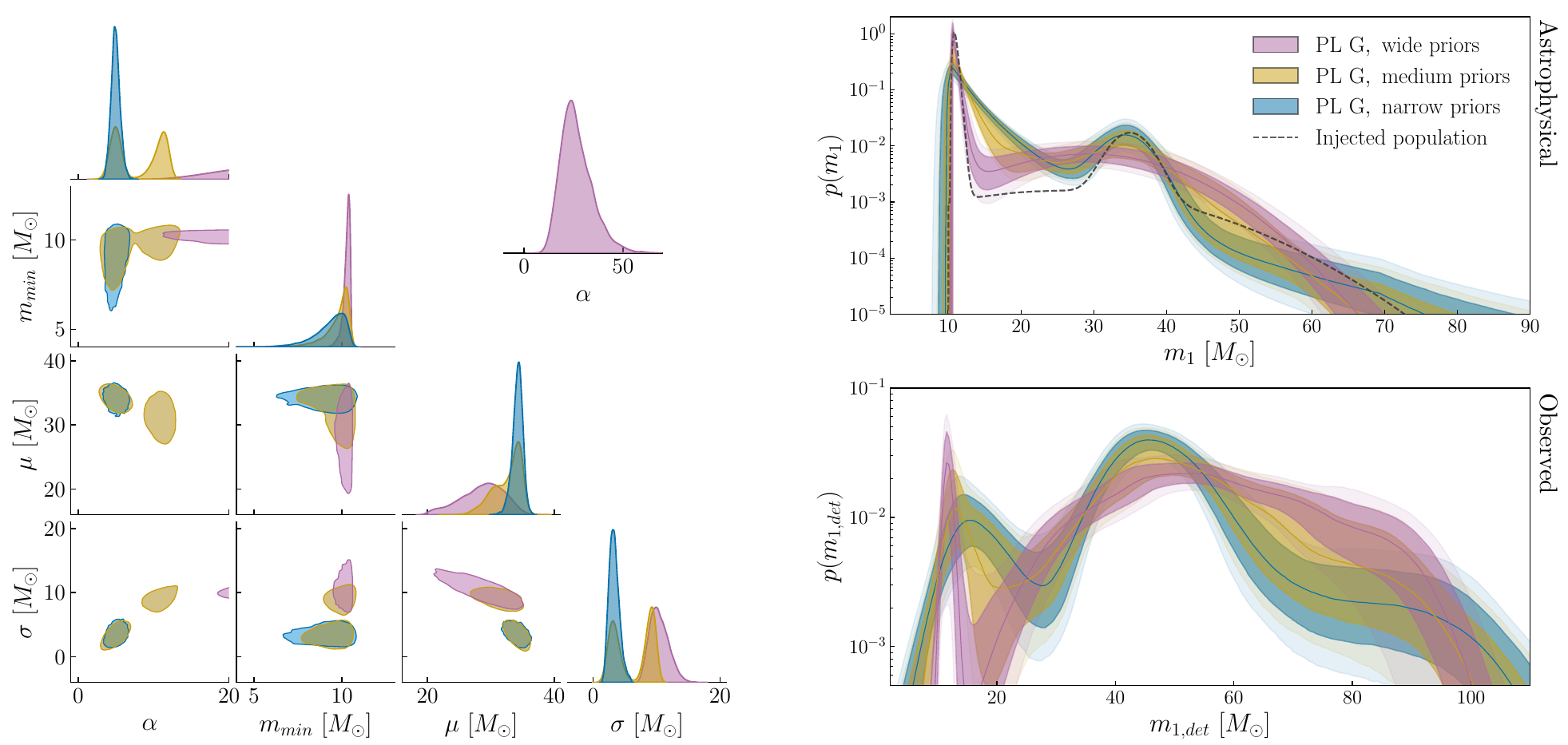}
    \caption{\justifying
    Comparison between different priors on the Powerlaw slope $\alpha$ and the Gaussian width $\sigma$ with the fiducial \model{Powerlaw-Gaussian} \footnotesize{model on \textbf{simulated data}. On the \textit{left}, a corner plot of these parameters together with the Powerlaw minimum $m_{min}$ and the Gaussian mean $\mu$. For illustration purposes, the posteriors on $\alpha$ with wide priors are shown separately. On the \textit{right}, the median, $68\%, 90\%$ C.I. of the reconstructed astrophysical (\textit{top panel}) and observed (\textit{bottom panel}) primary mass distribution.}}
    \label{fig:O3_stationary_injection}
\end{figure*}

\textit{Prior bounds on \model{Powerlaw-Gaussian} --}
%
We now expand on the discussion in Sec.~\ref{Sec:stationary_models} for the prior bounds with the \model{Powerlaw-Gaussian} model.
Following the hypothesis that the data contain more features than the model used, we draw events from an injected population that has both a sharp Powerlaw peak at $\sim 10 M_{\odot}$ and a Gaussian overdensity at $\sim 35 M_{\odot}$ (see the grey dashed line in Fig.~\ref{fig:O3_stationary_injection}).
For this purpose, we use a (stationary) \model{Powerlaw-Gaussian-Gaussian} model with the following injected population values
\begin{align*}
    &\alpha = 50, m_{\text{min}} = 10, m_{\text{max}} = 100, \delta_m = 3,\\
    &\mu_{a} = 25, \sigma_{a} = 15, \mu_{b} = 35, \sigma_{b} = 2.5,\\
    &\text{mix}_{\alpha} = 0.85, \text{mix}_{\beta} = 0.05, \mu_{q} = 0.7,\\
    &\sigma_{q} = 0.1, \gamma = 3, \kappa = 0, z_p = 3.
\end{align*}
With this choice, we obtain a realization of $51$ detected events ($\text{SNR}\geq 12$), similar to the O3 dataset of $50$ events.
In recovery, we use a \model{Powerlaw-Gaussian} model with the following priors
\begin{itemize}
    \item \textit{narrow priors} $\quad \alpha \in (-4,8),\, \sigma \in (1,6)$,
    \item \textit{medium priors} $\quad \alpha \in (-4,12),\, \sigma \in (1,10)$,
    \item \textit{wide priors} $\quad \alpha \in (-4,100),\, \sigma \in (1,30)$,
\end{itemize}
while those on the other parameters are listed in Tab.~\ref{Tab:models_priors}.

Although we are able to reproduce the results on O3 data, we note that the injected population is not necessarily the real astrophysical one.
In fact, as discussed in Sec.~\ref{Sec:stationary_models}, the \model{Powerlaw-Gaussian-Gaussian} model used in this simulation is not the preferred one on the data.
Finally, note that in the posteriors in Fig.~\ref{fig:O3_stationary_injection} we do not show the injected values because the model used in the recovery is different from the one used to generate the events.

\section{Additional results with O3 data}\label{App:O3_additional_results}

In this section we report other relevant results obtained by analyzing the O3 data.

In general, since most of our results support sharp features such as the $\sim 10 M_{\odot}$ peak, it is important to ensure that the likelihood calculation is numerically stable~\cite{Talbot:2023pex}.
We quantify this effect using numerical stability estimators that estimate the number of points used to compute the Monte Carlo integrals, the \textit{effective number of posterior samples} per event $N_{\text{eff,PE}}$ and the \textit{effective number of injections} $N_{\text{eff,INJ}}$ (Eqs. 4-8 of~\cite{Mastrogiovanni:2023zbw}).
Following~\cite{Farr:2019rap, KAGRA:2021duu}, we reject population samples for which $N_{\text{eff,PE}}<10$ and $N_{\text{eff,INJ}} < 4\,N_{\text{events}} = 200$.
Below, we report some values of $N_{\text{eff,PE}}$ and $N_{\text{eff,INJ}}/4\,N_{\text{events}}$ obtained for the maximum likelihood model of some relevant analyses,
\small{
\begin{itemize}
    \item \model{Powerlaw-Gaussian} narrow priors\\
    $\quad N_{\text{eff,PE}} = 92.3,\, N_{\text{eff,INJ}}/4\,N_{\text{events}} = 70.0$,
    \item \model{Powerlaw-Gaussian} wide priors\\
    $\quad N_{\text{eff,PE}} = 34.3,\, N_{\text{eff,INJ}}/4\,N_{\text{events}} = 21.0$,
    \item \model{Powerlaw-Powerlaw-Powerlaw}\\
    $\quad N_{\text{eff,PE}} = 12.8,\, N_{\text{eff,INJ}}/4\,N_{\text{events}} = 7.2$,
    \item \model{Powerlaw-Powerlaw-Powerlaw} redshift evolving\\
    $\quad N_{\text{eff,PE}} = 24.9,\, N_{\text{eff,INJ}}/4\,N_{\text{events}} = 13.6$.
\end{itemize}
}\normalsize{}
The values decrease as more sharp features are added to the model, and only the $N_{\text{eff,PE}}$ for the \model{Powerlaw-Powerlaw-Powerlaw} model is relatively close to the stability threshold.
To make sure that our results are numerically stable, we perform the stationary \model{Powerlaw-Powerlaw-Powerlaw} analysis with twice the threshold for both the effective number of posterior samples and injections, i.e. $N_{\text{eff,PE}}>20$ and $N_{\text{eff,INJ}}>400$.
The results obtained are identical to the default settings.
Additionally, we have also tested the wide priors \model{Powerlaw-Gaussian} model without any cut in the likelihood, again obtaining the same results.
\\

\begin{figure*}
    \hspace*{-0.2cm}\includegraphics[scale=0.9]{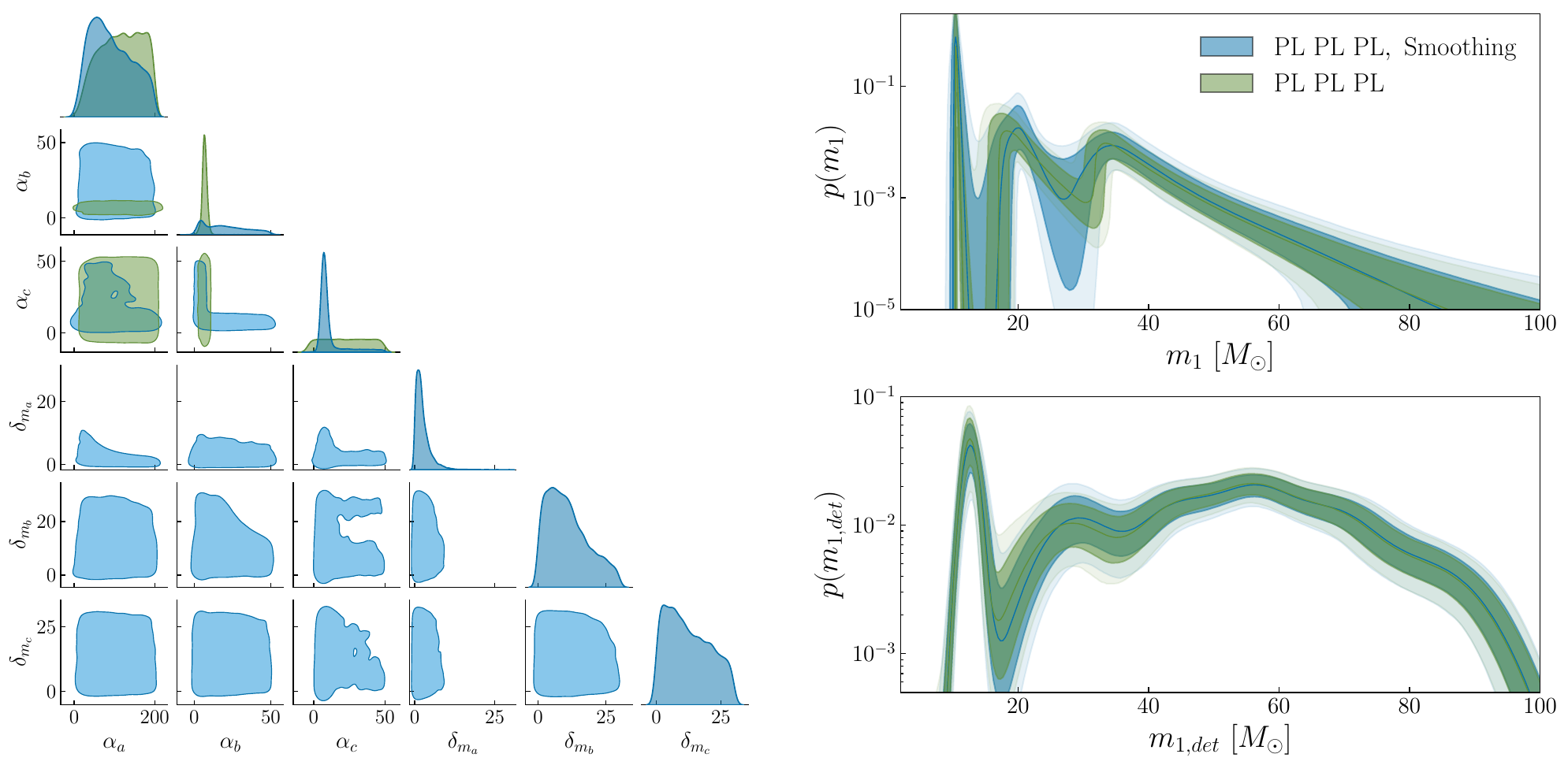}
    \caption{\justifying
    Comparison of O3 data results obtained with a stationary \model{Powerlaw-Powerlaw-Powerlaw} \footnotesize{model with (blue) and without (green) applying a window function on the Powerlaws. On the \textit{left}, a corner plot of the Powerlaw spectral indices $\alpha_i$ and the smoothing widths $\delta_{m_i}$. On the \textit{right}, the median, $68\%, 90\%$ C.I. of the reconstructed astrophysical (\textit{top panel}) and observed (\textit{bottom panel}) primary mass distribution.}}
    \label{fig:O3_PL3-smoothing}
\end{figure*}

\textit{Powerlaw smoothing --}
We briefly describe how the results with the stationary \model{Powerlaw-Powerlaw-Powerlaw} model are affected by the presence of the smoothing function.
To prevent the mass distribution from having unphysical sharp edges, it is common to apply a windowing function to the minimum mass truncation of the Powerlaw.
We use the sigmoid-like window function defined in Eqs. B11-B12 in~\cite{Mastrogiovanni:2023zbw}, whose width is regulated by the population parameter $\delta_m$.

In the \model{Powerlaw-Gaussian} analysis in Sec.~\ref{Sec:stationary_models}, the posteriors on $\delta_m$ change from being broadly peaked at $\sim 4 M_{\odot}$ in the case of ``narrow priors'', to railing against zero with ``wide priors''.
This behaviour is a consequence of the sharp $\sim 10 M_{\odot}$ peak, whose sharp profile is inconsistent with a larger smoothed peak.
To test the impact of the window functions on the \model{Powerlaw-Powerlaw-Powerlaw} model, we compare the posteriors and reconstructed distributions in Fig.~\ref{fig:O3_PL3-smoothing} with and without smoothing.
From the corner plot on the left, we first note that the $\delta_{m_a}$ posterior is zero, consistent with the \model{Powerlaw-Gaussian} model, while the $\delta_{m_b}$ and $\delta_{m_c}$ posteriors are mostly uninformative and only slightly bent towards zero.
This uncertainty is reflected in the reconstruction of the astrophysical mass spectrum (top right), which shows a larger uncertainty around the underdensity regions.
Interestingly, the introduction of the window functions reverses the role of the second and third Powerlaw slopes, $\alpha_b$ and $\alpha_c$, the former being well measured without smoothing and becoming uninformative with smoothing, and vice versa for $\alpha_c$.
This effect changes the mass spectrum reconstruction around $\sim 25 M_{\odot}$ to such an extent that the observed distributions are almost in disagreement at a $68\%$ credible interval.
This interesting phenomenology, which points to the limitations of the \model{Powerlaw-Powerlaw-Powerlaw} model, is reinforced by the limited information in the data, further supported by a negative BF for the analysis with the applied smoothing, $\text{ln}\,\mathcal{B} = -1.9 \pm 0.2$.
For this reason, we mainly consider unsmoothed models, both in the stationary and in the redshift-evolving case.

Finally, we briefly comment on the slope of the $\sim 10 M_{\odot}$ peak.
Without applying any window function, the posteriors on $\alpha_a$ are in general relatively uninformative, only disfavouring low spectral indices.
This effect is related to the fact that the resolved peak is so sharp that large changes in the spectral index values give very similar mass spectra.
This points out to possible limitations of this parametrization, which is not flexible enough to properly capture all the information at $\sim 10 M_{\odot}$ in the data, as discussed in Sec.~\ref{Sec:stationary_models}.
However, the inclusion of smoothing seems to mitigate this effect, allowing a broad peak in the $\alpha_a$ posterior to be better resolved, as shown in Fig.~\ref{fig:O3_PL3-smoothing}.\\

\begin{figure}[H]
    \hspace*{-0.4cm}\includegraphics[scale=0.5]{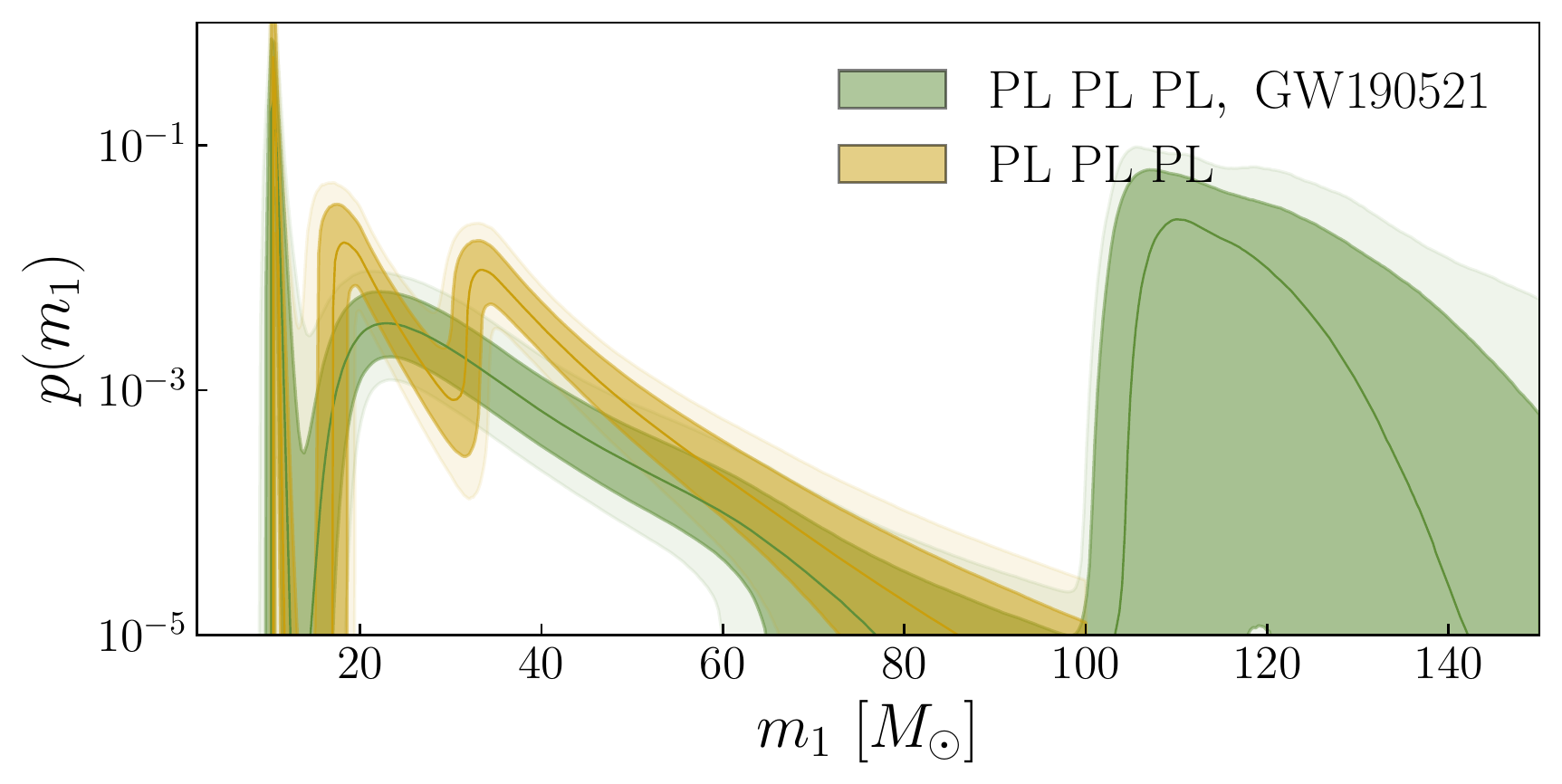}
    \caption{\justifying
    Comparison of O3 data results obtained with a stationary \model{Powerlaw-Powerlaw-Powerlaw} \footnotesize{model with (green) and without (yellow) including the event \GW{190521}. The figure shows the median, $68\%, 90\%$ C.I. of the reconstructed astrophysical primary mass distribution.}}
    \label{fig:GW190521_PL3}
\end{figure}

\textit{Including \GW{190521} --}
In our analysis, we exclude the high mass event \GW{190521}~\cite{LIGOScientific:2020iuh, LIGOScientific:2020ufj}.
Due to its peculiar mass falling in the predicted mass gap produced by pair instability in supernovae, this event has been intensively studied to understand whether it is consistent with the rest of the primary mass distribution or whether it represents a population outlier~\cite{LIGOScientific:2020iuh, Fishbach:2020qag, Nitz:2020mga, Edelman:2021fik}.
In our analysis, we find a curious result when using pure Powerlaw models with this event.
Given that \GW{190521} is relatively separated from the rest of the primary distribution, we observe that the high-mass Powerlaw exclusively fits the event instead of the rest of the distribution.
Fig.~\ref{fig:GW190521_PL3} shows such behaviour for the \model{Powerlaw-Powerlaw-Powerlaw} model, although similar conclusions were initially obtained with a \model{Powerlaw-Powerlaw}.
To avoid such problematic behaviour, we decided to remove \GW{190521} from the dataset.

We mention that the results in Sec.~\ref{Sec:stationary_models} on the prior bounds are not affected by the presence of \GW{190521}, and the same effect can be reproduced with slightly different choices of the priors on $\sigma$, i.e. $\sigma \in (1, 10)$.\\

\textit{Evolving maximum mass --}
The redshift evolving analysis in Sec.~\ref{Sec:O3_evolving_models} assumes that the maximum mass of the Powerlaws is stationary, i.e. $m_{\text{max}_{z_1}} = 0$.
This simplification has been made to reduce the number of free parameters in the analysis, especially given the fact that the maximum mass is generally poorly measured even in the stationary case (see the bottom-right corner in Fig.~\ref{fig:O3_PL3_corners}).
Nonetheless, we test the possibility of an evolving maximum mass using a \model{Powerlaw-Powerlaw-Powerlaw} model in which we let all the population parameters of the third Powerlaw, ``c'', evolve linearly with redshift.
The resulting posterior distributions are shown in the bottom-right corner of Fig.~\ref{fig:O3_PL3_corners}, where all the coefficients in the redshift expansion in Eq.~\eqref{eq:linear_expansion} are given for the minimum and maximum mass of the Powerlaw ``c''.
The posteriors on $m_{\text{max}}$ are still largely uninformative, and not much information can be obtained from the data.

\section{Dataset}\label{sec:dataset}

In this work, we use the GW signals from the third LVK observing run~\cite{LIGOScientific:2020ibl, LIGOScientific:2021djp}, O3, which are listed in Tab.~\ref{Tab:events}.
These correspond to the set of events with $\text{IFAR}>4$.
Additionally, two events are excluded from the analysis.
\GW{190412} is a highly asymmetric binary with a relatively small mass ratio that is outside the rest of the population distribution~\cite{LIGOScientific:2020stg, Rinaldi:2023bbd}.
For this reason, we decided not to consider it as it could impact the results by not following the mass ratio parametrization used, i.e. the truncated \model{Gaussian}.
We also do not consider the high-mass event \GW{190521}, for reasons that have been discussed in App.~\ref{App:O3_additional_results}.

We decide not to include the O1 and O2 events in our study.
This choice is motivated by the fact that no real-noise injections selected by $\text{FAR}$ are currently available at O1 and O2 sensitivities, as the LVK injection sets only provide semianalytic sensitivity estimates based on a prediction of the optimal SNR~\cite{O1O2O3:bbhpop}.
Since we only use $\text{IFAR}$ for the detection criterion, including O1-O2 events in our dataset would introduce bias in our analysis, whereas using O3-only injections~\cite{O3:bbhpop} and O3 data makes our analysis consistent.
Additionally, we select the events using a more stringent cut of $\text{IFAR}>4$ compared to the one used by the LVK study on the GWTC-3 population of $\text{IFAR}>1$ for BHs.
This reduces the contamination from false triggers, at the cost of further limiting our dataset.
The LVK study on cosmology~\cite{LIGOScientific:2021aug} uses an additional $\text{SNR}$ cut of 11 over the same $\text{IFAR}$ cut we adopt.
As a result, our choice provides us with a dataset ``between'' the LVK population and cosmology analyses.

\begin{figure}[h!]
    \hspace*{-0.15cm}\includegraphics[scale=0.5]{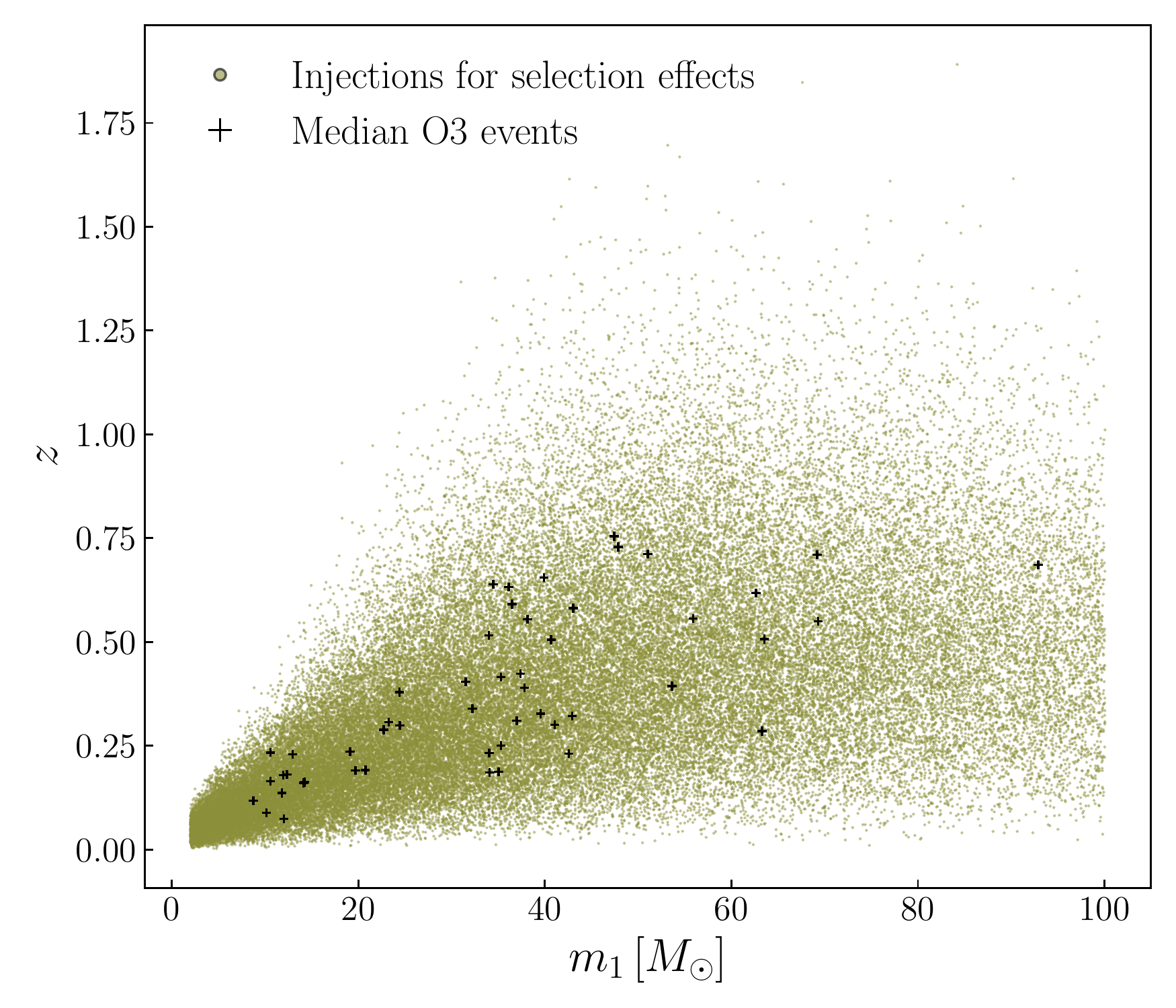}
    \caption{\justifying \footnotesize{Median value of the O3 events posteriors (\textit{black}) and injections used to compute selection effects (\textit{acid green}).
    The high mass event is \GW{190521}.
    The injections are filtered with an $\text{IFAR}>4$.}}
    \label{fig:O3_events_scatter}
\end{figure}

\begin{center}
    \begin{tabular}{p{2.7cm}p{2.7cm}p{2.4cm}}
        \hline
        \bf{events} \\
        \hline
        \\[-3mm]
        \GW{190408\_181802} & \GW{190720\_000836} & \GW{191215\_223052} \\
        \GW{190413\_134308} & \GW{190727\_060333} & \GW{191216\_213338} \\
        \GW{190421\_213856} & \GW{190728\_064510} & \GW{191222\_033537} \\
        \GW{190503\_185404} & \GW{190803\_022701} & \GW{191230\_180458} \\
        \GW{190512\_180714} & \GW{190828\_063405} & \GW{200112\_155838} \\
        \GW{190513\_205428} & \GW{190828\_065509} & \GW{200128\_022011} \\
        \GW{190517\_055101} & \GW{190910\_112807} & \GW{200129\_065458} \\
        \GW{190519\_153544} & \GW{190915\_235702} & \GW{200202\_154313} \\
        \GW{190521\_074359} & \GW{190924\_021846} & \GW{200208\_130117} \\
        \GW{190527\_092055} & \GW{190925\_232845} & \GW{200209\_085452} \\
        \GW{190602\_175927} & \GW{190929\_012149} & \GW{200219\_094415} \\
        \GW{190620\_030421} & \GW{190930\_133541} & \GW{200224\_222234} \\
        \GW{190630\_185205} & \GW{191105\_143521} & \GW{200225\_060421} \\
        \GW{190701\_203306} & \GW{191109\_010717} & \GW{200302\_015811} \\
        \GW{190706\_222641} & \GW{191127\_050227} & \GW{200311\_115853} \\
        \GW{190707\_093326} & \GW{191129\_134029} & \GW{200316\_215756} \\
        \GW{190708\_232457} & \GW{191204\_171526} & \\
        \hline
    \end{tabular}
    \captionof{table}{\footnotesize Table with the O3 events used in the analysis.}\label{Tab:events}
\end{center}


\begin{figure*}
    \hspace*{-0.2cm}\includegraphics[scale=0.92]{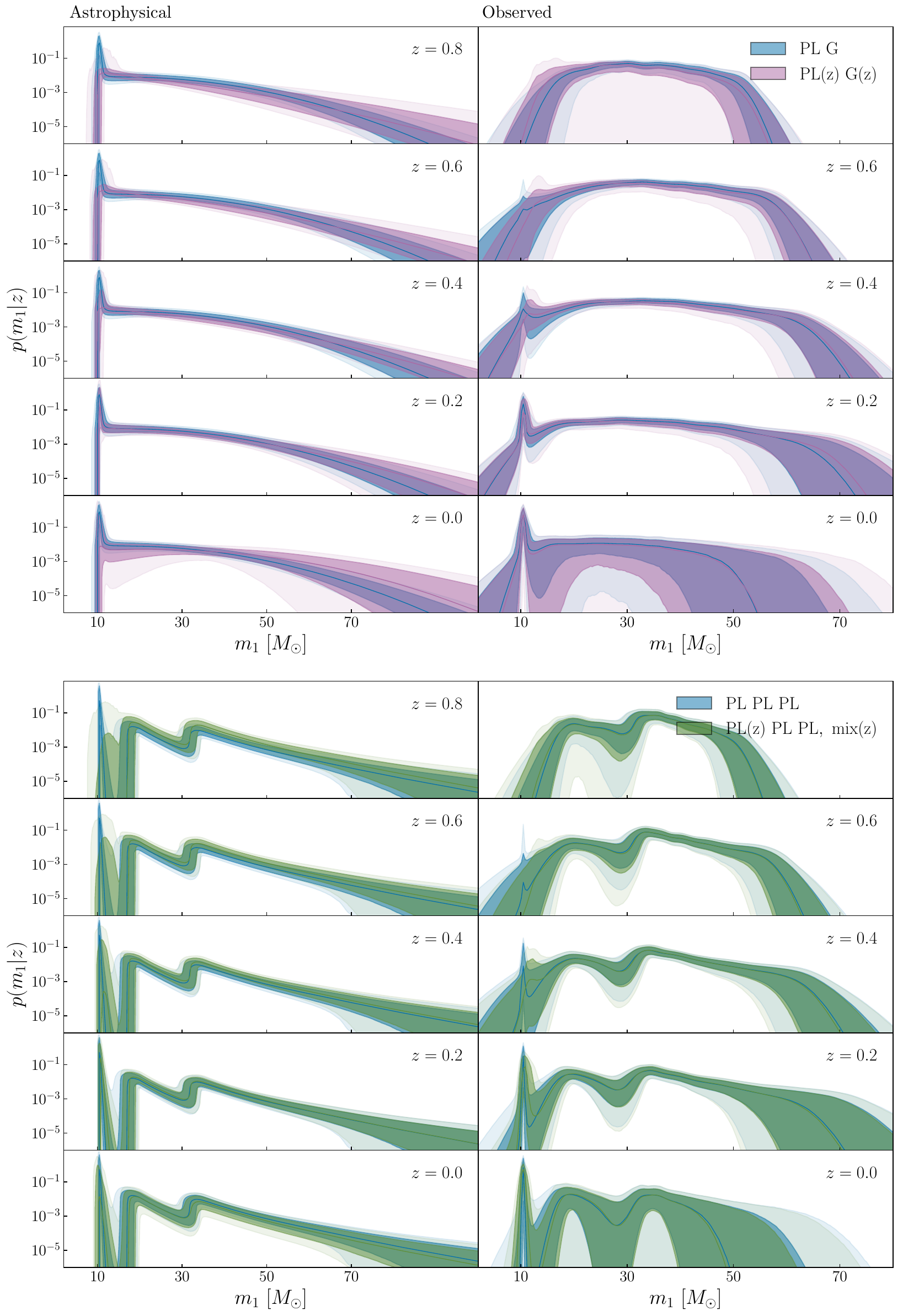}
    \caption{\justifying \footnotesize{Comparison on O3 data of the astrophysical (\textit{left column}) and observed (\textit{right column}) reconstructed primary distributions as a function of redshift between the stationary (blue) and redshift evolving (pink and green) distributions. \textit{Top panel}:} \model{Powerlaw-Gaussian} \footnotesize{model with the two features evolving linearly in redshift and the mixture function assumed stationary. \textit{Bottom panel:}} \model{Powerlaw-Powerlaw-Powerlaw} \footnotesize{model with only the first Powerlaw evolving linearly in redshift and a linearly evolving mixture function. The countours mark the the median, $68\%, 90\%$ C.I.}
    }
    \label{fig:O3_evolving_PL-G-PL3-transition}
\end{figure*}

\begin{figure*}
    \hspace*{-0.22cm}\includegraphics[scale=0.61]{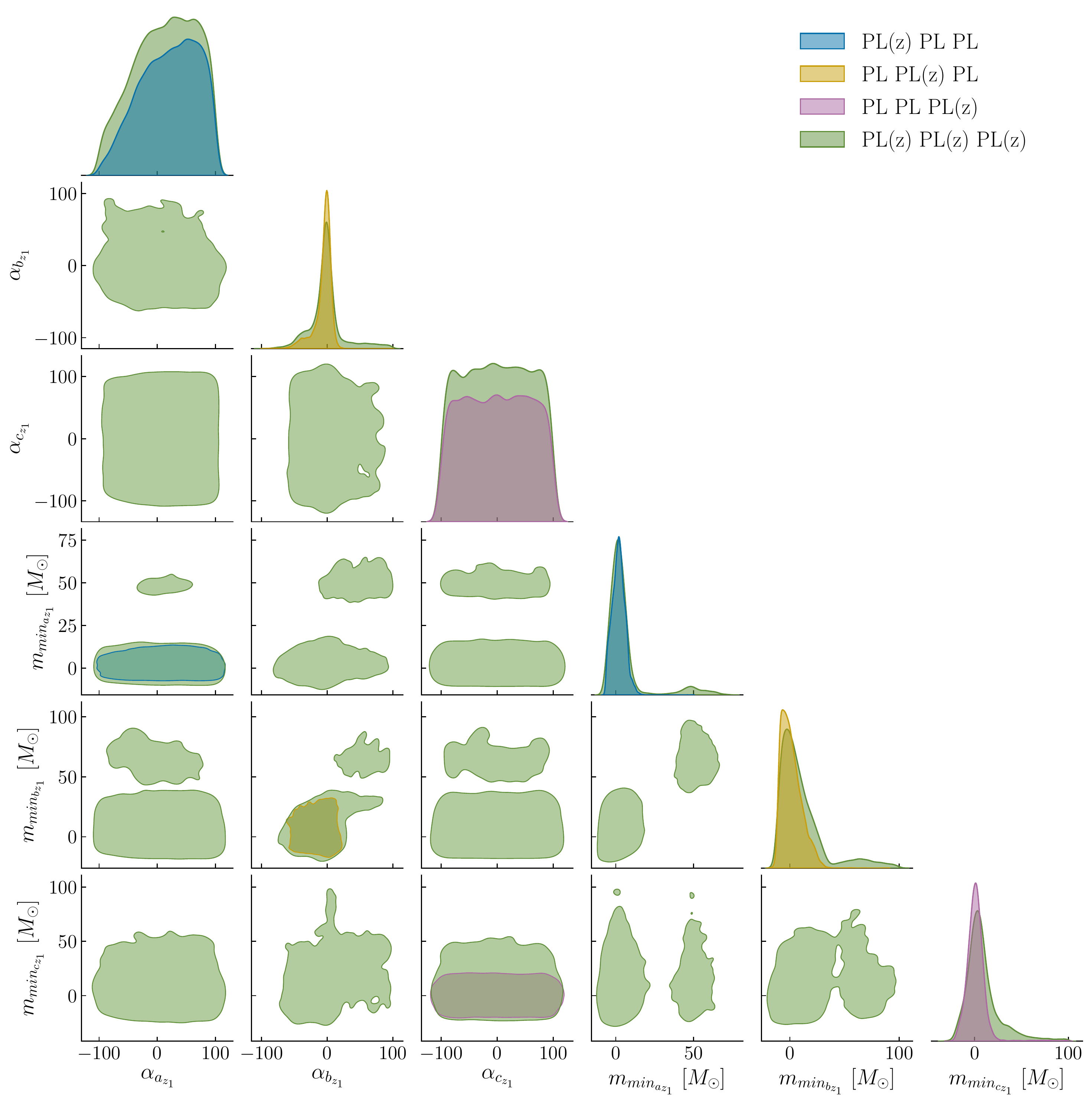} 
    \caption{\justifying \footnotesize{For the O3 data, corner plot with the posterior distributions of the linear redshift expansion coefficients in Eq.~\eqref{eq:linear_expansion}, for different combinations of evolving features using the} \model{Powerlaw-Powerlaw-Powerlaw} \footnotesize{model. The analyses assume a stationary mixture function.}}
    \label{fig:O3_evolving_3-features}
\end{figure*}

\begin{figure*}
    \hspace*{-0.22cm}\includegraphics[scale=0.61]{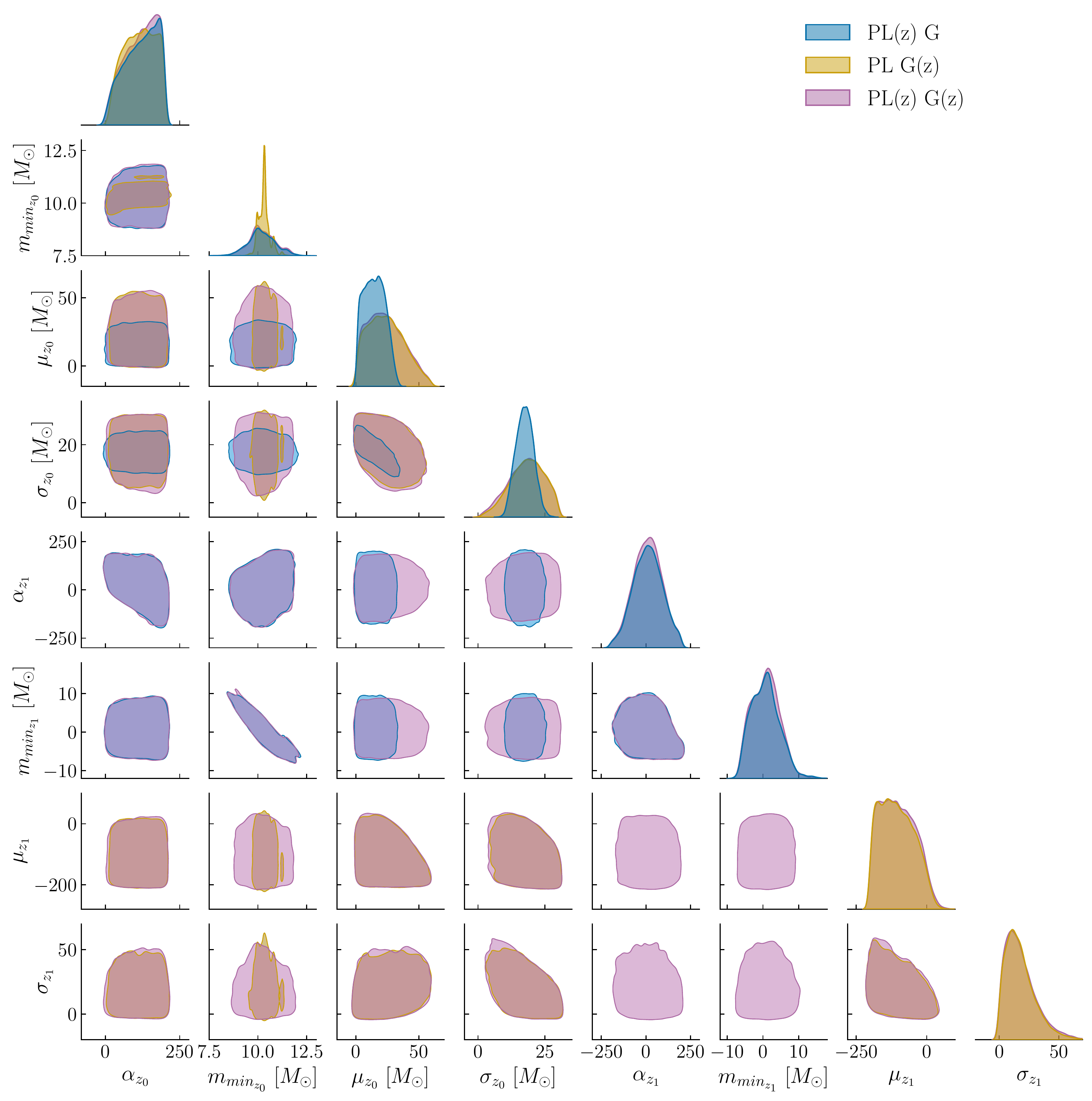} 
    \caption{\justifying \footnotesize{For O3 data, corner plot with the posterior distributions of the stationary and linear redshift expansion coefficients in Eq.~\ref{eq:linear_expansion}, for different combinations of evolving features using the} \model{Powerlaw-Gaussian} \footnotesize{model. The analyses assume a stationary mixture function.}}
    \label{fig:O3_evolving_2-features}
\end{figure*}

\begin{figure*}
    \hspace*{-0.22cm}\includegraphics[scale=0.9]{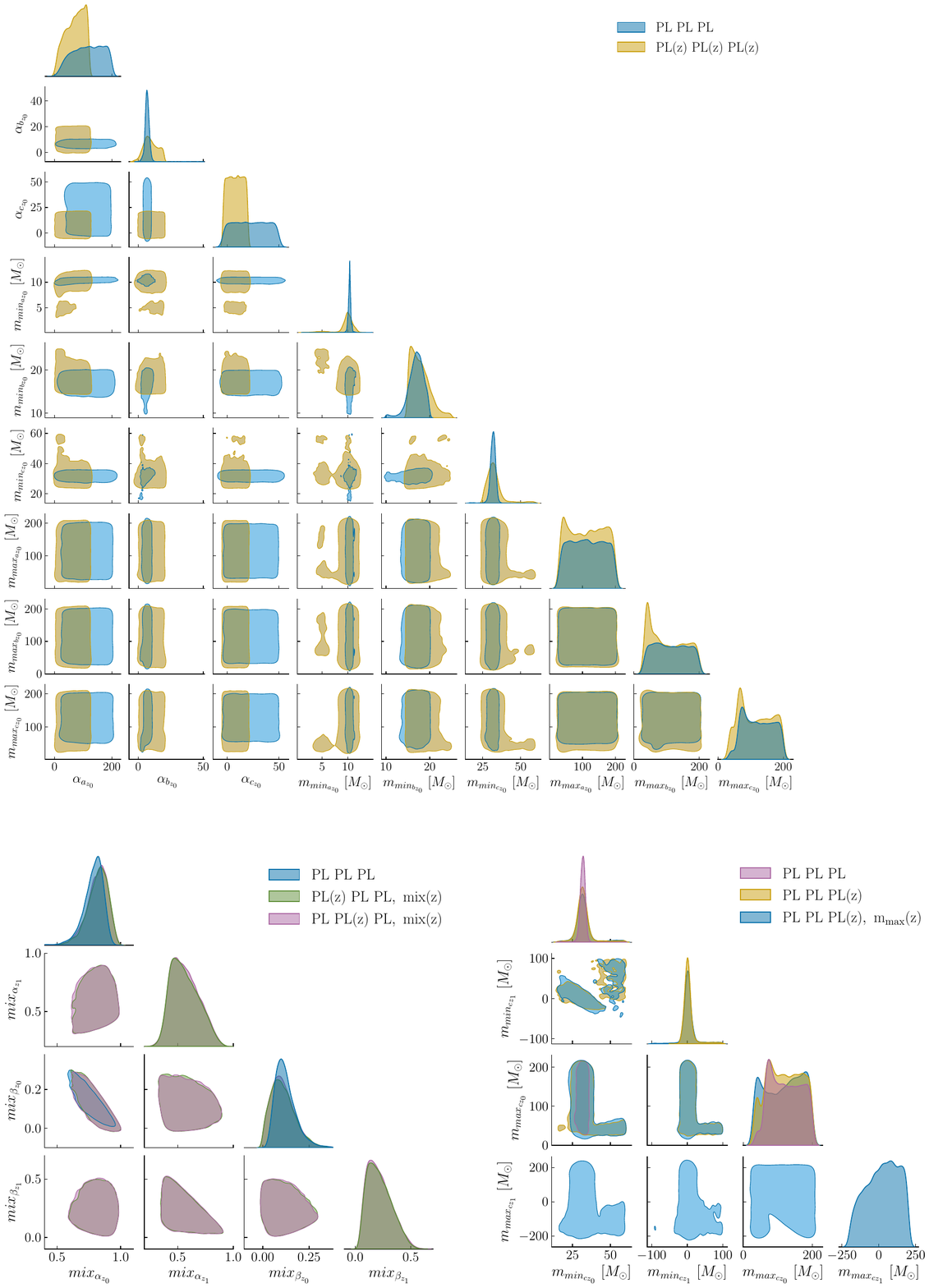}
    \caption{\justifying \footnotesize{For O3 data, corner plots of different population parameters and parametrizations specified in the legend, using the \model{Powerlaw-Powerlaw-Powerlaw} \footnotesize{model. On \textit{top}, a comparison between the stationary (blue) and redshift evolving (yellow) case on the common non-evolving parameters. \textit{Bottom left}, posteriors on the redshift evolving mixture functions. \textit{Bottom right}, posteriors on the minimum and maximum mass of the third Powerlaw, including the case of a redshift evolving maximum mass.}}}
    \label{fig:O3_PL3_corners}
\end{figure*}


\clearpage
\bibliography{references}

\end{document}